\documentstyle[12pt,amsmath,amssymb,epic]{article}

\newcommand{\bbbone}{\mathchoice {\rm 1\mskip-4mu l} {\rm 1\mskip-4mu l}
{\rm 1\mskip-4.5mu l} {\rm 1\mskip-5mu l}}

\newcommand{\scalprod}[2]{\left\langle {#1}, {#2}\right\rangle}
\newcommand{\dom}{{\cal D}}
\newcommand{\RE}{{\rm Re}}
\newcommand{\IM}{{\rm Im}}
\newcommand{\fer}[1]{(\ref{#1})}
\newcommand{\ran}{{\rm Ran\,}}
\newcommand{\pomegaperp}{P_{\Omega}^\perp}

\newcommand{\ppperp}{P_{\Omega^p_\beta}^\perp}
\newcommand{\repsilonbar}{\overline{\!R}_\epsilon}
\newcommand{\repsilon}{R_\epsilon}
\newcommand{\qbar}{\overline{Q}}

\newcommand{\edeltanot}{E_\Delta^0}
\newcommand{\pfinot}{P_{\Omega_{\beta,0}}}
\newcommand{\pfinotperp}{\pfinot^\perp}
\newcommand{\h}{{\cal H}}
\newcommand{\norm}[1]{\left|\!\left| #1 \right|\!\right|}
\newcommand{\lless}{<\!\!<}
\newcommand{\ggeq}{>\!\!>}
\newcommand{\Span}{{\rm span}}
\newcommand{\Arctan}{{\rm Arctan}}
\newcommand{\supp}{{\rm supp}}
\newcommand{\edprime}{F_{\Delta'}^0}

\newcommand{\nr}{{\cal N}_r}
\newcommand{\nl}{{\cal N}_l}
\newcommand{\tr}{{\mbox{tr\,}}}

\newcommand{\av}[1]{\left\langle{#1}\right\rangle}

\begin{document}

\title{Positive Commutators in Non-Equilibrium Quantum Statistical Mechanics
\thanks{This work is part of the  author's PhD requirement.}}
\author{Marco Merkli\\
Department of Mathematics, University of Toronto
\thanks{present address: Department of Mathematics, ETH Z\"urich, merkli@math.ethz.ch}
}
\date{May 8, 2001}
\maketitle

\begin{abstract}
The method of positive commutators, developed for zero temperature problems over the last twenty years, has been an essential tool in the spectral analysis of Hamiltonians in quantum mechanics. We extend this method to positive temperatures, i.e. to non-equilibrium quantum statistical mechanics.\\
\indent
We use the positive commutator technique to give an alternative proof of a fundamental property of a certain class of large quantum systems, called {\it Return to Equilibrium}. This property says that equilibrium states are (asymptotically) stable: if a system is slightly perturbed from its equilibrium state, then it converges back to that equilibrium state as time goes to infinity. 
\end{abstract}

\noindent
{\rm \small {\bf Keywords:}\ positive commutator, Mourre estimate, return to equilibrium, virial theorem, Fermi golden rule
}


\setcounter{section}{0}

\section{Introduction}

In this paper, we study a class of open quantum systems consisting of two interacting subsystems: a finite system, called the particle system coupled to a reservoir (heat bath), described by the spatially infinitely extended photon-field (a massless Bose field). The dynamics of the coupled system on the von Neumann algebra of observables is generated by a {\it Liouville operator}, also called Liouvillian or thermal Hamiltonian, acting on a positive temperature Hilbert space. Many key properties of the system, such as return to equilibrium (RTE), i.e. asymptotic stability of the equilibrium state, can be expressed in terms of the spectral characteristics of this operator.\\
\indent
Applying the positive commutator (PC) method to the Liouville operator of systems in question, we obtain rather detailed  information on the spectrum of these operators. This allows us to recover, with a partial improvement, a recent fundamental result by several authors on RTE.

Our main technical result is a positive commutator estimate (also called a
Mourre estimate) for the Liouville operator. This
result holds for a wider class of systems than previously considered.

Spectral information on the Liouville operator, and hence the property of RTE,  is extracted from the PC estimate through Virial Theorem type arguments. It turns out that the existing Virial Theorem techniques are too restrictive to apply to positive temperature systems, and we need to extend them beyond their traditional range of application.

There is a restriction on the class of systems for which we prove RTE, due to our Virial Theorem type result mentioned above. This is
the first result of this kind, and we expect that it will be improved to yield
the RTE result for a considerably wider class of systems.

\subsection{A class of open quantum systems}

The choice of the class of systems we analyze is motivated by the quantum mechanical models of 
nonrelativistic matter coupled to the radiation field, or matter interacting with a phonon field (quantized modes of a
lattice), or a generalized spin-boson system. For notational convenience, we consider only scalar Bosons. A good review
of physical models leading to the class of Hamiltonians considered here is found in
[HSp].\\

{\bf The non-interacting system.\  }
The algebra of observables of the uncoupled system is the $C^*$-algebra ${\frak A}={\cal B}({\cal H}_p)\otimes{\cal W}({\frak H}_0)$, where ${\cal B}({\cal H}_p)$ denotes the bounded operators on the particle Hilbert space ${\cal H}_p$ and ${\cal W}({\frak H}_0)$ is the Weyl CCR algebra  over the one-particle space ${\frak H}_0=\{f\in L^2({\mathbb R}^3, d^3k): \int |k|^{-1}|f(k)|^2<\infty\}$. The restriction to $f\in {\frak H}_0$ comes from the fact that we will work in the Araki-Woods representation of the CCR algebra, which is only defined for Weyl operators $W(f)$ with $f\in{\frak H}_0$ (see [AW], [JP1,2], [BFS4]). The dynamics of the non-interacting system is given by the automorphism group ${\mathbb R}\ni t\mapsto \alpha_{t,0}\in\mbox{Aut}({\frak A})$, $\alpha_{t,0}(A)=e^{itH_0}Ae^{-itH_0}$, where $H_0=H_p\otimes \bbbone_f+\bbbone_p\otimes H_f$ is the sum of the particle and free field Hamiltonians. $H_0$ acts on the Hilbert space ${\cal H}_p\otimes {\cal H}_f$, where ${\cal H}_f=\bigoplus_{n=0}^\infty {\frak H}_0^{\otimes^n_{\rm sym}}$ is the Fock space over ${\frak H}_0$ and $H_f$ is the free field Hamiltonian, i.e. the second quantization of the multiplication operator by $\omega=|k|$, $H_f=\mbox{d}\Gamma(\omega)$; if $a^*(k), a(k)$ denote the (distribution valued) creation and annihilation operators, then we can express it equivalently as $H_f=\int \omega(k)a^*(k)a(k)d^3k$. The particle Hamiltonian is assumed to be a selfadjoint operator  on ${\cal H}_p$ which has purely discrete spectrum:
\begin{equation}
\sigma(H_p)=\{E_j\}_{j=0}^\infty,
\label{cond1}
\end{equation}
(where multiplicities are included, i.e. for a degenerate eigenvalue $E_i$, we have $E_i=E_j$ for some $j\neq i$), 
and we denote the orthonormal basis diagonalizing $H_p$ by $\{\varphi_j\}$. Let $\mbox{tr}$ denote the trace on ${\cal B}({\cal H}_p)$, then we further assume that 
\begin{equation}
Z_p(\beta):=\tr e^{-\beta H_p}<\infty,\ \ \ \forall \beta>0.
\label{trace}
\end{equation}
We do not need to further specify the particle system. As
a concrete example, one may think of a system of finitely many Schr\"odinger
particles in a box (hence the name particle system), or a spin system. In some of our results (see Theorem 4.4 on the Fermi Golden Rule Condition), we shall assume that the spectrum of $H_p$ is finite ($N$-level system).\\
\indent
The equilibrium state at temperature $T=1/\beta>0$ for the non-interacting system is given by the product $\omega_{\beta,0}=\omega_\beta^p\otimes\omega_\beta^f\in{\frak A}^*$. Here, $\omega^p_\beta(\cdot)=\frac{\tr(e^{-\beta H_p}\ \cdot)}{\tr e^{-\beta H_p}}$ is the particle-Gibbs state at temperature $\beta$ and $\omega^f_\beta$ is the field $\beta$-KMS state that describes the infinitely extended field in the state of black body radiation, i.e. its two-point function is given according to Planck's law by  $\omega^f_\beta(a^*(k)a(k'))=\frac{\delta(k-k')}{e^{\beta|k|}-1}$. The GNS construction for $({\cal A},\alpha_{t,0}, \,\omega_{\beta,0})$ yields the (up to unitary equivalence) unique data $({\cal H},L_0,\Omega_{\beta,0},\pi)$ (dependent on $\beta$). Here, ${\cal H}$ is the GNS Hilbert space with inner product $\scalprod{\cdot}{\cdot}$, $\Omega_{\beta,0}$ is a cyclic vector for the $*$-morphism $\pi:{\frak A}\rightarrow {\cal B}({\cal H})$ (the representation map), and {\it the Liouvillian} $L_0$ is the selfadjoint operator on ${\cal H}$ implementing the dynamics, i.e. satisfying $L_0\Omega_{\beta,0}=0$ and 
\begin{equation*}
\omega_{\beta,0}(\alpha_{t,0}(A))=\scalprod{\Omega_{\beta,0}}{e^{itL_0}\pi(A)e^{-itL_0}\Omega_{\beta,0}}, \ \ \ \forall A\in{\frak A}.
\end{equation*}
This GNS construction has been carried out in [AW] (for the field, the particle part is standard since it is a finite system), see also [JP1,2], [BFS4]. We shall not explicitly use the representation map $\pi$ here and thus omit its presentation which can be found in the above references. The GNS Hilbert space and cyclic vector are given by
\begin{eqnarray}
{\cal H}&=&{\cal H}_p\otimes{\cal H}_p\otimes{\cal F}(L^2({\mathbb R}\times S^2)),\label{PTHS}\\
\Omega_{\beta,0}&=&\Omega_\beta^p\otimes\Omega,\label{XXp}
\end{eqnarray}
where  $\Omega_\beta^p$ is the particle Gibbs state at temperature $\beta$ given in \fer{particlegibbs}. ${\cal F}(L^2({\mathbb R}\times S^2))$ is the Fock space over $L^2({\mathbb R}\times S^2)$ with vacuum $\Omega$, which we call the {\it Jak\u si\'c-Pillet glued space}. It was introduced by Jak\u si\'c and Pillet in [JP1] and is isomorphic to ${\cal H}_f\otimes{\cal H}_f$, the field GNS Hilbert space constructed in [AW]. It is easily verified that the Liouvillian is given by $L_0=L_p+L_f$ (see also [JP1,2]). We write simply $L_p$ instead of $L_p\otimes\bbbone_{{\cal F}(L^2({\mathbb R}\times S^2))}$ and similarly for $L_f$. Here, $L_p=H_p\otimes\bbbone_p-\bbbone_p\otimes H_p$, $L_f=\mbox{d}\Gamma(u)$ and $u$ is the first (the radial) variable in ${\mathbb R}\times S^2$. It is clear that the spectrum of $L_p$ is the  discrete set $\{e=E_i-E_j: E_{i,j}\in\sigma(H_p)\}$ and the spectrum of $L_f$ is the entire real axis (continuous spectrum) with an embedded eigenvalue at $0$ (corresponding to the vacuum eigenvector $\Omega$). Consequently, $L_0$ has continuous spectrum covering the whole real line and embedded eigenvalues given by the eigenvalues of $L_p$.\\

{\bf The interacting system.\ }
We now describe the interacting system by defining an interacting Hamiltonian acting on ${\cal H}_p\otimes{\cal H}_f$: 
\begin{equation}
H=H_0+\lambda v,
\label{H}
\end{equation}
where the coupling constant $\lambda$ is a small real number, and   
\begin{equation}
v=G\otimes(a(g)+a^*(g)).
\label{v}
\end{equation}
Here, $G$ is a bounded selfadjoint operator on ${\cal H}_p$. The function
$g\in{\frak H}_0$ is called the {\it form factor} and the smoothed out creator is given by
$
a^*(g)=\int d^3k\, g(k)a^*(k).
$
We assume $g$ to be a bounded
$C^1$-function, satisfying the following infra-red (IR) and ultra-violet (UV)
conditions (recall that $\omega=|k|$):
\begin{equation}
\begin{array}{ll}
\mbox{IR:} & \mbox{$|g(k)|\leq C\omega^p$, for some $p>0$, as
  $\omega\rightarrow 0$},\\
&\mbox{\ for some results, we assume $p>2$},\\
\mbox{UV:} & \mbox{$|g(k)|\leq C\omega^{-q}$, for some $q>5/2$, as
  $\omega\rightarrow\infty$}.
\end{array}
\label{IRUV}
\end{equation}
In addition, we assume that conditions \fer{IRUV} hold for the
  derivative $\partial_\omega g$, if $p,q$ are replaced by $p-1, q+1$.\\
\indent
We point out that the value coming from the model of an atom coupled to the radiation field in the dipole approximation is $p=1/2$ (without this approximation, $p=-1/2$). From now on we will refer to $p=1/2$ as the physical case.\\
\indent
The interacting Hamiltonian (which describes the coupled system at zero temperature) corresponds to an interacting Liouvillian (positive temperature Hamiltonian) which is given by (c.f. [JP1,2], [BFS4]):
\begin{eqnarray}
L&=&L_0+\lambda I,\label{ell}\\
I&=&G_l\otimes\left(a^*(g_1)+a(g_1)\right)-G_r\otimes \left(a^*(g_2)+a(g_2)\right).
\label{017}
\end{eqnarray}
Here, $G_l:=G\otimes\bbbone_p$, $G_r:=\bbbone_p\otimes{\cal C}G{\cal C}$, where ${\cal C}$ is the antilinear map on ${\cal H}_p$ that, in the basis that diagonalizes $H_p$, has the effect of complex conjugation of coordinates. The origin of ${\cal C}$ is the identification of the Hilbert-Schmidt operators on ${\cal H}_p$ with ${\cal H}_p\otimes{\cal H}_p$ via the isomorphism $|\varphi\rangle\langle\psi|\leftrightarrow \varphi\otimes{\cal C}\psi$ (see also [JP2], [BFS4]). Moreover, we have defined, for $g\in L^2({\mathbb R}_+\times S^2)$:
\begin{equation}
g_1(u,\alpha)=
 \left\{
  \begin{array}{ll}
   \sqrt{1+\mu(u)}\  u \ g(u,\alpha), & u\geq 0\\
    \sqrt{\mu(-u)} \ u\   \overline{g}(-u,\alpha), & u<0
   \end{array}
  \right.
\label{i)}
\end{equation}
and  $g_2(u,\alpha)=-g_1(-u,\alpha)$, where the function $\mu=\mu(k)$ is the momentum density distribution, given by Planck's law describing black body radiation:
$
\mu(k)=(e^{\beta\omega}-1)^{-1}$, $\omega=|k|$. The structure of $g_1$ in \fer{i)} comes from the Jak\u si\'c-Pillet gluing which identifies $L^2({\mathbb R}^3)\oplus L^2({\mathbb R}^3)$ with $L^2({\mathbb R}_+\times S^1)$ via the isometric isomorphism $(f_1,f_2)\mapsto f$, $f(u,\alpha)=uf_1(u,\alpha)$ for $u\geq 0$ and $f(u,\alpha)=u\overline{f}_2(-u,\alpha)$ for $u<0$. For more detail, we refer to [JP1,2].\\
\indent
For $\lambda\neq 0$, one can construct a vector $\Omega_{\beta,\lambda}\in {\cal H}$ s.t. the state defined by $\omega_{\beta,\lambda}(A)=\scalprod{\Omega_{\beta,\lambda}}{A\Omega_{\beta,\lambda}}$ is a $\beta$-KMS state w.r.t. the coupled dynamics $\alpha_{t}(A)=e^{itL}Ae^{-itL}$, where $A$ is an element in the von Neumann algebra ${\frak M}:={\cal B}({\cal H}_p)\otimes \overline{\pi({\cal W}({\frak H}_0))}$ (weak closure in ${\cal B}({\cal F}(L^2({\mathbb R}\times S^2)))$\,). An extension of the algebra of observables to this weak closure is necessary since the full dynamics does not leave ${\cal B}({\cal H}_p)\otimes\pi({\cal W}({\frak H}_0))$ invariant. It is not difficult to show that $({\frak M},\alpha_t)$ is a $W^*$-dynamical system (compare also to [FNV], [JP2]). Notice in particular that $L\Omega_{\beta,\lambda}=0$.\\
\indent
The construction of $\Omega_{\beta,\lambda}$ goes under the name {\it structural stability of KMS states}, see [BFS4] for this specific model, but also [A], [FNV], [BRII]. For $\beta|\lambda|$ small, one has the estimate (for the $O$-notation, see after \fer{27}):
\begin{equation}
\|\Omega_{\beta,\lambda}-\Omega_{\beta,0}\|=O(\beta|\lambda|).
\label{pertkms}
\end{equation}
We show in Appendix A.1 that $L$ is essentially selfadjoint (Theorem A.2).

\subsection{Spectral characterization of RTE}
We define the  equilibrium states at temperature $T=1/\beta>0$ to be the $\beta$-KMS states. Hence the equilibrium state of the coupled system at inverse temperature $\beta>0$ is given by the above constructed $\omega_{\beta,\lambda}\in{\frak M}^*$. A conjectured property of KMS states is their {\it dynamical stability} (which should be a natural property of equilibrium states). In our case, this means that $\omega'\circ\alpha_t\rightarrow\omega_{\beta,\lambda}$ as $t\rightarrow\infty$, for states $\omega'$ that are close to $\omega_{\beta,\lambda}$. This is called the property of return to equilibrium. Apart from specifying the mode of convergence, it remains to say what we mean by $\omega'$ is close to $\omega_{\beta,\lambda}$. 
There is a natural neighbourhood of states around $\omega_{\beta,\lambda}$ in which the dynamics is also determined by $L$: the set of all {\it normal} states $\omega'$ w.r.t. $\omega_{\beta,\lambda}$.  By definition, $\omega'$ is normal w.r.t. $\omega_{\beta,\lambda}$, iff 
\begin{equation}
\forall A\in{\frak M}:\ \  \omega'(A)=\tr(\rho A),
\label{close}
\end{equation}
where $\tr(\cdot)$ is the trace on the GNS Hilbert space ${\cal H}$ given in \fer{PTHS}, $\rho$ is a trace class operator on ${\cal H}$, normalized as $\tr\rho=1$.\\

{\bf Proposition 1.1 (spectral characterization of RTE).\ }{\it Let ${\frak M}\subset {\cal B}({\cal H})$ be a von Neumann algebra and suppose that $\omega_\beta(\cdot)=\scalprod{\Omega_\beta}{\cdot\,\Omega_\beta}: {\frak M}\rightarrow {\mathbb C}$ is a $\beta$-KMS state with respect to the dynamics $\alpha_t\in \mbox{Aut}({\frak M})$.  Suppose that the Liouvillian $L$ generating the dynamics on
  ${\cal H}$ has no eigenvalues except for a simple one at zero, so that the
  only eigenvector of $L$ is $\Omega_\beta$. Then, for any normal state
  $\omega'$ w.r.t. $\omega_\beta$, and for any observable $A\in{\frak M}$, we have 
\begin{equation}
\lim_{T\rightarrow\infty}\frac{1}{T}\int_0^T\omega'(\alpha_t(A))dt=\omega_\beta(A).
\label{a15}
\end{equation}
This means that the system exhibits return to equilibrium in an ergodic mean sense.
}\\

The proof is given e.g. in [JP2], [BFS4], [M].  Better information on the spectrum of $L$ yields stronger convergence; if $L$ has absolutely continuous spectrum, except a simple eigenvalue at $0$, then \fer{a15} can be replaced by $\lim_{t\rightarrow\infty}\omega'(\alpha_t(A))=\omega_\beta(A)$.

\subsection{The PC method}
This section introduces the general idea of the PC method. As we have seen above, the Liouville operators in the class of systems we consider consist of two parts: 
\begin{equation*}
L=L_0+\lambda I,
\end{equation*}
where $L_0$ is the uncoupled Liouville operator, describing the two subsystems
(particles and field) when they do not interact. $I$ is the interaction, and
$\lambda$ is a real (small) coupling parameter. 
The spectrum of $L_0$ consists of a
continuum covering the whole real axis, and it has embedded eigenvalues, arranged symmetrically w.r.t. zero. Moreover, zero is a
degenerate eigenvalue. 
We would like to show that for
$\lambda\neq 0$, the spectrum of $L$ has no eigenvalues, except for a simple
one at zero, because then Proposition 1.1 tells us that the system exhibits RTE!\\
\centerline{\ }

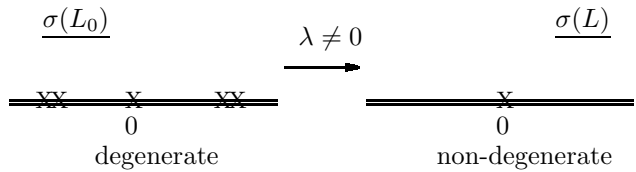
\centerline{\input{thesfig.epic}}

\centerline{\ }
\noindent
In other words, we want to show that all nonzero eigenvalues of $L_0$ are
unstable under the perturbation $\lambda I$, and that this perturbation removes the degeneracy
of the zero eigenvalue. We know that  $L$ has a zero eigenvalue with eigenvector $\Omega_{\beta,\lambda}$, the perturbed KMS state. This means that our task reduces to
showing instability of all nonzero eigenvalues, and that the dimension of the
nullspace of $L$ is at most one.\\

It is conventional wisdom that embedded eigenvalues are unstable under generic
perturbations,  turning into resonances. We now outline the technique
we use to show instability of embedded eigenvalues: the PC technique.\\
\indent
To do so, we concentrate first on a nonzero (isolated) eigenvalue $e$ of $L_0$
whose instability we want to show. The main idea is to construct an anti-selfajoint operator $A$, called the {\it adjoint operator} (to $L$), s.t. we
have the following PC estimate:
\begin{equation}
E_\Delta(L)[L,A]E_\Delta(L)\geq\theta E_\Delta^2(L),
\label{intro1}
\end{equation}
where $\theta>0$ is a srictly positive number, $E_\Delta(L)$ denotes the
spectral projector of $L$ onto the interval $\Delta$, and $[\cdot,\cdot]$ is
the commutator. Here, $\Delta$ is chosen to contain
the eigenvalue $e$ but no other eigenvalues of $L_0$. Equation \fer{intro1} is
also called a (strict) {\it Mourre estimate}. If it is satisfied, then one
 sees that $L$ has no eigenvalues in $\Delta$ by using the
following argument by contradiction: suppose that $L\psi=e'\psi$, with $e'\in\Delta$  and
$\|\psi\|=1$. Then we have $E_\Delta(L)\psi=\psi$, and the PC estimate
\fer{intro1} gives on one hand
$
\scalprod{\psi}{[L,A]\psi}\geq\theta.
$
On the other hand,
formally expanding the commutator yields
\begin{equation}
\scalprod{\psi}{[L,A]\psi}=\scalprod{\psi}{[L-e',A]\psi}=2\RE\scalprod{(L-e')\psi}{A\psi}=0,
\label{intro2}
\end{equation}
which leads to the contradiction $\theta\leq 0$, hence showing that there cannot
be any eigenvalue of $L$ in $\Delta$.\\
\indent
This formal proof is in general wrong. Indeed, both
operators $L$ and $A$ are unbounded, and one has to take great care of domain
questions, including the very definition of the commutator $[L,A]$.\\
\indent
Relation \fer{intro2} is called the {\it Virial Theorem}, and it can be made
in many concrete cases rigorous by approximating the hypothetical
eigenfunction $\psi$ by ``nice'' vectors. The situation in which this works is
quite generally given by the case where $[L,A]$ is bounded relatively to $L$, which is in particular satisfied for $N$-body Schr\"odinger
systems, and systems of particles coupled to a field {\it at zero
  temperature}. However, in our case the condition is not satisfied, and as mentioned above, we
have to develop a more general argument of this type.\\
\indent
The treatment of the zero eigenvalue is similar, except that we prove \fer{intro1} only on $\ran E_\Delta(L)P^\perp$, where $P$
is the rank-one projector onto the known zero eigenvector $\Omega_{\beta,\lambda}$ of $L$, and $P^\perp$ is its orthogonal complement.

\section{Main results}

Our main technical result is the abstract PC estimate, Theorem 2.1. This
result is the basis for the spectral analysis of the Liouvillian, as explained above. We point out that the PC estimate holds for infra-red behaviour of the form factor (see \fer{IRUV}) characterized by $p>0$, which
covers the physical case $p=1/2$.\\
\indent
Theorem 2.2 characterizes the spectrum of the Liouvillian in view of the
property of RTE. To
prove this result, we combine the PC estimate with a Virial Theorem type
argument. It is for the latter that we need presently the more restricting
infra-red behaviour $p>2$. We think that our method can be improved.\\
\indent
A direct consequence of Theorem 2.2 is Corollary 2.3 which says that the system exhibits RTE (recall also Proposition 1.1).\\
\indent
All the results hold under assumption of the Fermi Golden Rule Condition, \fer{fgrc} and \fer{fgrc0}. In Theorem 2.4, we give explicit conditions on the operator $G$ and the form factor $g$ so that the {\it Fermi Golden Rule Condition} holds. We start by explaining this condition. In the language of quantum resonances, it expresses the fact that the bifurcation of complex eigenvalues (resonance poles) of the spectrally deformed Liouvillian takes place at second order in the perturbation (i.e. the lifetime of the resonance is of the order $\lambda^{-2}$).\\
\indent
As we have mentioned above, the Liouvillian corresponding to the particle system at positive
temperature is given by $L_p=H_p\otimes\bbbone-\bbbone\otimes H_p$, acting on the Hilbert space ${\cal H}_p\otimes{\cal H}_p$, so $L_p$  has discrete spectrum given by
$
\sigma(L_p)=\{e=E_i-E_j: E_i,E_j\in\sigma(H_p)\}.
$
For every eigenvalue $e$ of $L_p$, we define an operator $\Gamma(e)$ acting on
the corresponding eigenspace, $\ran P(L_p=e)\subset {\cal H}_p\otimes {\cal H}_p$, by 
\begin{equation}
\Gamma(e)=\int_{{\mathbb R}\times S^2} m^*(u,\alpha)P(L_p\neq
e)\delta(L_p-e+u) m(u,\alpha),
\label{Gamma}
\end{equation}
where $\delta$ denotes the Dirac function, and where the operator $m$ is given
by 
\begin{equation}
m(u,\alpha)=G_l \, g_1(u,\alpha) -G_r \, g_2(u,\alpha).
\label{m}
\end{equation}
Recall that $g_{1,2}$ and $G_{l,r}$ were defined in and before equation \fer{i)}.\\
\indent
It is clear from \fer{Gamma} that $\Gamma(e)$
is a non-negative selfadjoint operator. The Fermi Golden Rule Condition is used to show instability of
embedded eigenvalues. For nonzero eigenvalues, the condition says
that $\Gamma(e)$ is {\it strictly positive}:
\begin{equation}
\mbox{for $e\neq 0$,\ \ \ \ } \gamma_e:=\inf\sigma\left(\Gamma(e)\upharpoonright
\ran P(L_p=e)\right)>0.
\label{fgrc}
\end{equation}
\indent
We show in Theorem 2.4 that $\Gamma(0)$ has a simple eigenvalue at zero, the eigenvector
being the Gibbs state of the particle system, $\Omega^p_\beta$ (see \fer{particlegibbs}). This
reflects the fact that the zero eigenvalue of $L_0$ survives the  
perturbation, however, its degeneracy is removed, i.e. the zero eigenvalue of
$L$ is simple. The Fermi Golden Rule Condition for $e=0$ requires strict positivity  on the complement
of the zero eigenspace of $\Gamma(0)$, i.e.
\begin{equation}
\gamma_0:=\inf\sigma\left(\Gamma(0)\upharpoonright
\ran P(L_p=0)P^\perp_{\Omega^p_\beta}\right)>0.
\label{fgrc0}
\end{equation}
Here, $P_{\Omega^p_\beta}$ is the projection onto ${\mathbb C}\Omega^p_\beta$, and $P^\perp_{\Omega_\beta^p}=\bbbone-P_{\Omega_\beta^p}$. 
We give in Theorem 2.4 below  explicit conditions on $G$ and $g(k)$ s.t. \fer{fgrc}
and \fer{fgrc0} hold.\\

\noindent
Here is our main result.\\

{\bf Theorem 2.1 (Positive Commutator Estimate).\ }{\it Assume the IR and UV
  behaviour \fer{IRUV}, with $p>0$. Let $\Delta$ be an interval containing
  exactly one eigenvalue $e$ of $L_0$ and let $h\in C^\infty_0$ be a smooth function s.t. $h=1$ on
  $\Delta$ and $\supp h\cap\sigma(L_p)=\{e\}$. Assume the Fermi Golden Rule Condition
  \fer{fgrc} (or \fer{fgrc0}) holds. Let $\beta\geq\beta_0$, for any fixed $0<\beta_0<\infty$. Then there is a $\lambda_0>0$ (depending on $\beta_0$) s.t. if $0<|\lambda|<\lambda_0$, then we have in the sense of quadratic forms on $\dom(N^{1/2})$ (see remark 1 below), for some explicitely constructed anti-selfadjoint operator $A$:
\begin{equation}
h(L) [L,A]h(L)\geq {\textstyle{\frac{1}{2}}} \lambda^{91/50} h(L) \left(\gamma_e(1-5\delta_{e,0}P_{\Omega_{\beta,0}})-O(\lambda^{1/200})\right)h(L).
\label{27}
\end{equation}
}

{\it Notation.\ } Let $s$ be a real variable. Then $O(s)$ stands for a family $T_s$ of bounded operators depending on $s$, satisfying $\lim_{s\rightarrow 0}\|T_s\|/s=C<\infty$. In \fer{27}, $s=\lambda^{1/200}$.\\
\indent  
{\it Remarks.\ } 
1.\, $N=\mbox{d}\Gamma(1)$ is the number operator in the positive temperature Hilbert space (see also 
\fer{PTHS} and \fer{nop}), and $P_{\Omega_{\beta,0}}$ is the projector onto the span of  $\Omega_{\beta,0}$, the $\beta$-KMS state of
the uncoupled system (see \fer{XXp}). Also,  $\delta_{e,0}$ is the Kronecker
symbol, equal to one if $e=0$ and zero else.\\
2. \, We show in Theorem A.2 that $L$ is
essentially selfadjoint on a dense domain in the positive temperature Hilbert space.\\
3.\, The commutator $[L,A]$ is by construction in first approximation equal to
$N$ (see Section 7), and $h(L)$ leaves the domain $\dom(N^{1/2})$ invariant (see e.g. [M]), so that  \fer{27} is well defined.\\
4.\, There is no smallness condition on the interval $\Delta$ (apart from it only containing one eigenvalue of $L_0$).\\

{\bf Theorem 2.2 (Spectrum of $L$).\ }{\it Assume the IR condition $p>2$ (see
  \fer{IRUV}). Let $\beta\geq\beta_0$, for any fixed $0<\beta_0<\infty$, $\beta<\infty$. Then the Liouvillian $L$ has the following spectral properties:
\begin{itemize}
\item[1)] Let $e\neq 0$ be a nonzero eigenvalue of $L_0$, and suppose that the
  Fermi Golden Rule Condition \fer{fgrc} holds for $e$. Then there is a $\lambda_0>0$ (dependent on $\beta_0$) s.t. for $0<|\lambda|<\lambda_0$, $L$ has no eigenvalues in the open interval $(e_-,e_+)$, where $e_-$ is
the biggest eigenvalue of $L_0$ smaller than $e$, and $e_+$ is the smallest
eigenvalue of $L_0$ bigger than $e$.
\item[2)] Assume the Fermi Golden Rule Condition \fer{fgrc0}  holds for $e=0$. Then there is a $\lambda_0>0$ (dependent on $\beta_0$) s.t. if $0<|\lambda|<\lambda_0$ and $0<\beta|\lambda|<\lambda_0$, then  $L$ has a simple
  eigenvalue at zero.\\
\end{itemize}
}

{\it Remark.\ } Theorem 2.2 shows that if the Fermi Golden Rule Condition holds for all
  eigenvalues of $L_0$, then $L$ has no eigenvalues, except a simple one
  at zero.\\

{\bf Corollary 2.3 (Return to Equilibrium).\ }{\it Suppose the IR condition and the condition on $\beta$ as
  in Theorem 4.2, and that the Fermi Golden Rule Condition is satisfied for
  all eigenvalues of $L_0$. If $|\lambda|>0$ is small (in the sense of Theorem 4.2, 2), then every normal state w.r.t. the $\beta$-KMS
  state 
  $\Omega_{\beta,\lambda}$ (the zero eigenvector of $L$) exhibits return to
  equilibrium in an ergodic mean sense.}\\

The Corollary follows immediately from Theorem 2.2 and
Proposition 1.1, where the ergodic mean convergence is defined by \fer{a15}.\\

{\bf Theorem 2.4 (Spectrum of $\Gamma(e)$).\ }{\it Set $\Gamma_p(e):=P(L_p=e)\Gamma(e)P(L_p=e)$.  
\begin{itemize}
\item[{\bf 1)}] Let $e\neq 0$. Then there is a non-negative number $\delta_0=\delta_0(G)$ (independent of $\beta,\lambda$) whose value is given in Appendix A.2 (see before \fer{110}) s.t.
\begin{equation*}
\Gamma_p(e)\geq \delta_0 \inf_{\{E_{ij}\neq 0\}}\left(|E_{ij}|\int_{S^2}dS(\omega,\alpha) \left| g(|E_{ij}|,\alpha)\right|^2\right)  P(L_p=e).
\end{equation*}
In particular, the Fermi Golden Rule Condition \fer{fgrc} is satisfied if the
r.h.s. is not zero.
\item[{\bf 2)}]  $\Gamma_p(0)$  has an eigenvalue at
  zero, with the particle Gibbs state $\Omega^p_\beta$ as eigenvector:
\begin{equation}
\Omega^p_{\beta}=Z_p(\beta)^{-1/2}\sum_{i}e^{-\beta E_i/2}\varphi_i\otimes\varphi_i,
\label{particlegibbs}
\end{equation}
where we recall that $Z_p(\beta)$ was defined in \fer{trace}. Moreover, if
\begin{equation*}
g_0:=\inf_{\{E_{mn}<0\}}\left|\scalprod{\varphi_n}{G\varphi_m}\right|^2\frac{e^{\beta E_n}}{e^{-\beta E_{mn}}-1}\int_{{\mathbb R}^3}\delta(E_{mn}+\omega) |g|^2\geq0
\end{equation*}
is strictly positive, then zero is a simple eigenvalue of $\Gamma_p(0)$ with
unique eigenvector $\Omega^p_\beta$ and the spectrum of $\Gamma_p(0)$ has a gap
at\ zero: $(0,2g_0Z_p)\cap\sigma(\Gamma_p(0))=\emptyset$. In particular, the Fermi
Golden Rule Condition \fer{fgrc0} holds.\\
\end{itemize}
}
\noindent
{\it Remarks.\ } 1.\,  If $e\neq 0$ is nondegenerate, i.e. if $e=E_{m_0n_0}$ for a unique pair $(m_0,n_0)$, then (see before \fer{110})
$
\delta_0=\sum_{n\neq n_0}\left|\scalprod{\varphi_n}{G\varphi_{n_0}}\right|^2 +\sum_{m\neq m_0}\left|\scalprod{\varphi_m}{G\varphi_{m_0}}\right|^2.
$\\
2.\, If $H_p$ is unbounded, then $g_0=0$. Indeed, let $m$ be fixed, and take $n\rightarrow\infty$, then $E_{mn}<0$ and $\scalprod{\varphi_n}{G\varphi_m}\rightarrow 0$, since $\varphi_n$ goes weakly to zero. Notice though that $g_0>0$ is only a sufficient condition for the Fermi Golden Rule Condition to hold at zero.\\
3.\,  For $g_0>0$, the size of the gap, $2g_0Z_p$,  is bounded away from zero uniformly in $\beta\geq \beta_0$, since 
\begin{equation*}
\lim_{\beta\rightarrow\infty}\inf_{\{E_m<E_n\}}\frac{\tr e^{-\beta H_p}}{e^{-\beta E_m}-e^{-\beta E_n}} =\lim_{\beta\rightarrow\infty}\inf_{\{\hat{E}_m<\hat{E}_n\}}\frac{\tr e^{-\beta\hat{H}_p}}{e^{-\beta \hat{E}_m}-e^{-\beta \hat{E}_n}},
\end{equation*}
where  $\hat{E}_i:=E_i-E_0\geq 0$ ($E_0$ is the smallest eigenvalue of $H_p$)
and $\hat{H}_p:=H_p-E_0\geq 0$ (the smallest eigenvalue of $\hat{H_p}$ is
zero).

\section{Review of previous results}

Proving the RTE property is one of the key problems of non-equilibrium statistical mechanics. Until recently, this property was proven for specially designed abstract models (see [BRII]). The first result for realistic systems came in the pioneering work of Jak\u si\'c and Pillet [JP1,2] in 1996.\\
\indent
 In their work, Jak\u si\'c and Pillet prove return to equilibrium, with exponential rate of convergence in time, for the spin-boson system (i.e. an $N$-level system coupled to the free massless bosonic field with $N=2$; their work easily extends to general finite $N$) for sufficiently high temperatures. Their work introduces the spectral approach to RTE. The analysis is done in the spirit of the theory of quantum resonances, using {\it spectral deformation}
techniques, where the deformation is generated by energy-translation.
 The IR condition on the form factor is $g(\omega)\sim\omega^p$,
$\omega\rightarrow 0$, with $p>-1/2$, hence includes the physical case
$p=1/2$. However, there is a restriction on temperature:  $|\lambda|<1/\beta$. The spectral deformation technique imposes certain analyticity conditions on the form factor.\\
\indent
The $N$-level system coupled to the free massless bosonic field is also treated in
[BFS4], but the spectrum of the Liouvillian is analyzed using complex dilation
instead of translation. RTE with exponentially fast rate in convergence in
time is established for small coupling constant $\lambda$ {\it independent} of
$\beta$. Bach, Fr\"ohlich and Sigal adapt in this work their Renormalization
Group method developed in [BFS1,2,3] to the positive temperature case. The IR
condition is $p>0$, which includes the physical case.\\
\indent
In a recent work, Derezi\'nski and Jak\u si\'c [DJ] consider the Liouvillian
of the  $N$-level system interacting with the free massless bosonic
field. Their analysis of the spectrum of the Liouvillian is based on the {\it Feshbach method} which is justified with the help of the {\it Mourre Theory}, applied to the reduced Liouvillian (away from the vacuum sector). The Mourre theory in turn is based on a global positive commutator estimate for the reduced Liouvillian.
The IR condition for
instability of nonzero eigenvalues is $p>0$, and for the lifting of the
degeneracy of the zero eigenvalue, it is $p>1$.\\
\indent
The method for the spectral analysis of the Liouvillian we use employs the
energy-translation generator in the Jak\u si\'c-Pillet glued positive temperature
Hilbert space, as in [JP1,2] and [DJ]. We prove a Mourre estimate (PC estimate) for the original Liouvillian with a conjugate operator which is a deformation of the energy shift generator mentioned above. This method has been
developed in the zero-temperature case in [BFSS] (for the dilation generator
though).\\
\indent
Our construction of the PC works for the  IR
condition $p>0$, which includes the physical case. In order to conclude absence of eigenvalues from the PC estimate, the Virial
Theorem is needed. So far, the systems for which the Virial Theorem
was applied have always satisfied  
the condition that $[L,A]$ is relatively bounded with respect to $L$, in which
case a general theory has been developed, see [ABG] (for specific systems, see
also [BFSS] for particle-field at zero temperature, [HS1] for $N$-body
systems). We remark though that in [S], Skibsted extends the abstract Mourre theory to certain systems where $[L,A]$ is not relatively bounded (but $[[L,A],A]$ is). \\
\indent
We develop in this work a  Virial Theorem type argument in the case where the
commutator $[L,A]$ is not relatively $L$-bounded. This comes at the price
that our estimates involve the triple commutator $[[[L,A],A],A]$, and consequently, we need a restrictive IR behaviour of the form factor, namely 
$p>2$. We think that this restriction coming from
the part of the proof using the Virial Theorem (not the PC estimate), can be
improved by a better understanding of the Virial Theorem.\\
\indent
 It should be pointed out that the Virial Theorem is an important
tool of interest on its own, still currently under research, see e.g. [GG].\\

We finish this brief review by comparing our approach to that of [DJ] which, in the literature on the subject,  is closest to ours.\\
\indent
The main difference is that [DJ] develop first the Mourre theory for a {\it reduced} Liouville operator, staring from a global PC estimate on the radiation sector. Using the Feshbach method, they show then the limiting absorption principle for the Liouvillian acting on the full space. It is our impression that this method is restricted to systems where a global PC estimate is valid, i.e. for positive temperatures, one cannot avoid using the generator of translations as the adjoint operator.\\
\indent
The use of a different adjoint operator than the Jak\u si\'c-Pillet translation generator might be desirable, for instance in order to remove restrictive assumptions on the coupling functions.\\
\indent
In our method, we modify the bare adjoint operator in such a way as to have a local PC estimate right from the start for the {\it full} (i.e. not for a reduced) Liouvillian. This method has the advantage that it works for various choices of the adjoint operator, in fact, it was first developed (for zero temperatures) for the dilation generator in [BFSS]. It is true though that the use the translation generator greatly reduces the number of estimates to be performed, and this is the reason why we use it here.\\
\indent
Let us also mention that our PC estimate is local in the spectral localization of $L$, but is of a ``broad locality'' in the sense that it holds in neighbourhoods of eigenvalues of $L_0$ that need not be small, and are in particular independent of the coupling constant (the only restriction being that such neigbourhoods contain only one eigenvalue of the non-interacting Liouvillian). This means that we do not need two separate arguments to treat the regions ``close'' to (typically in a $\lambda^2$-neighbourhood) and ``away'' from the eigenvalues of $L_0$, as is often the case in Mourre theory, as well as in [DJ].\\
\indent
We do not claim that either of the two methods is better, both having, in our view, advantages and disadvantages. We do believe that our approach gives new insights and can open doors to new techniques to handle the problem of RTE and related spectral problems.

\section{Proof of Theorem 2.1: step 1.}

We prove in this section the PC estimate w.r.t. spectral localization in the
uncoupled Liouvillian $L_0$, see Theorem 4.3. Step 2 consists in passing from this estimate to
the one localized w.r.t. the full Liouvillian $L$ and is performed in the next section. \\
\indent
Our estimates are uniform in $\beta\geq\beta_0$ (for any $0<\beta_0<\infty$ fixed). For notational convenience, we set $\beta_0=1$, see also the remark after Proposition A.1 in Appendix A.1.

\subsection{PC with respect to spectral localization in $L_0$}

We construct an operator $B$ (see \fer{B}) which is positive on spectral subspaces of $L_0$, see Theorem 4.3 (the main result of this section).\\
\indent
On $L^2({\mathbb R}\times S^2)$ and for $t\in{\mathbb R}$, we define the unitary
transformation
$
\left(\tilde{U}_t\psi\right)(u,\alpha)=\psi(u-t,\alpha),
$
which induces a unitary transformation $U_t$ on Fock space ${\cal F}={\cal F}(L^2({\mathbb R}\times S^2))$:
$
U_t=\Gamma(\tilde{U}_t),
$
i.e. for $\psi\in{\cal F}$, the projection onto the $n$-sector of $U_t\psi$ is given by
$
\left(U_t\psi\right)_n(u_1,\ldots,u_n)=\psi_n(u_1-t,\ldots,u_n-t).
$ 
Here and often in the future, we do not display the angular variables $\alpha_1,\ldots,\alpha_n$ in  the argument
of $\psi_n$. $U_t$ is a strongly continuous unitary one-parameter ($t\in\mathbb
R$) group on ${\cal F}$. Its anti-selfadjoint generator $A_0$, defined in the strong sense by  $\partial_t|_{t=0}U_t=A_0$, is
$
A_0=-{\rm d}\Gamma(\partial_u).
$
The domain of the unbounded operator $A_0$, ${\cal D}(A_0)=\{\psi\in{\cal F}:
\partial_t|_{t=0}U_t\psi\in{\cal F}\}$, is dense in ${\cal F}$, which simply  follows from the fact that $A_0$
is the generator of a strongly continuous group. From now on, we write
$
U_t=e^{tA_0},\ t\in{\mathbb R}.
$
The following result serves to motivate the definition of an operator denoted by $[L,A_0]$ (see \fer{e1} below). The proof is not difficult and can be found in [M].\\

{\bf Proposition 4.1\ }{\it On the dense set $\dom(L_0)\cap\dom(N)$, we have
$
e^{-tA_0}Le^{tA_0}=L_0+tN+\lambda I_t,
$
where $I_t$ is obtained from $I$ by replacing the form factor $g$ by its
translate $g^t$, and $g^t(u,\alpha)=g(u+t,\alpha)$. We obtain therefore
\begin{equation}
\partial_t|_{t=0} e^{-tA_0}Le^{tA_0}=N+\lambda\tilde{I},
\label{018}
\end{equation}
where $\tilde{I}=G_l\otimes(a^*(\partial_ug_1)+a(\partial_ug_1))-G_r\otimes
(a^*(\partial_ug_2)+a(\partial_ug_2))$. The derivative in \fer{018} is
understood in the strong topology.}\\

On a formal level, we have
$
\partial_t|_{t=0}\ e^{-tA_0}Le^{tA_0}=-A_0L+LA_0=[L,A_0],
$
which suggests the {\it definition of the unbounded operator $[L,A_0]$} with domain $\dom([L,A_0])=\dom(N)$ as
\begin{equation}
\label{e1}
[L,A_0]:=N+\lambda \tilde{I}.
\end{equation}
We point out that the operator $[L,A_0]$ is defined as the r.h.s. of \fer{e1}, and
not as a commutator in the sense of $LA_0-A_0L$. 
Remark that $[L,A_0]$ is positive on $\dom(N)\cap\ran\pomegaperp$, where $\Omega$ is the vacuum in
${\cal F}$. Indeed, from Proposition A.1, it follows (take e.g. $c=1/4$)
$
[L,A_0]\geq\frac{3}{4} N-O(\lambda^2),
$
so that
$
\pomegaperp [L,A_0]\pomegaperp\geq\left(3/4-O(\lambda^2)\right)\pomegaperp.
$
On the other hand, $P_\Omega[L,A_0]P_\Omega=0$, so if we want to find an operator that
is positive also on ${\mathbb  C}\Omega$, then we need to modify $A_0$.\\ \indent For a fixed
eigenvalue $e\in\sigma(L_p)$, define
\begin{eqnarray}
b(e)&=&\theta\lambda \left(\qbar\repsilon^2IQ-QI\repsilon^2\qbar\right),\label{41.1}\\
R_\epsilon&=&\left((L_0-e)^2+\epsilon^2\right)^{-1/2}.\nonumber
\end{eqnarray}
Here, $\theta$ and $\epsilon$ are positive parameters, and $Q$, $\qbar$ are projection
operators on ${\cal H}$ defined as
\begin{equation}
Q=P(L_p=e) \otimes P_\Omega,\ \ \ \qbar=\bbbone-Q.
\label{Q}
\end{equation}
In what follows, we denote
$
\repsilonbar:=\qbar\repsilon.
$\\

{\bf Proposition 4.2\ }{\it The operator $b=b(e)$ is bounded and $[L,b]=Lb-bL$
  is well defined on ${\cal D}_0$ and it extends to a bounded operator on the
  whole space. We denote the extended operator again by $[L,b]$.}\\

{\it Proof.\ } The operator $b$ is bounded since both $IQ$ and $QI$ are bounded. Furthermore, since $||L_0\repsilon||\leq 1+|e|/\epsilon$ and $||L_0Q||=
|e|$, then $[L_0,b]$ is bounded. Moreover, since $||IQ||\leq C$ and
$||I\repsilonbar^2IQ||\leq C\epsilon^{-2}||(N+1)IQ||\leq 2C\epsilon^{-2}||IQ||\leq
C\epsilon^{-2}$, then also $||[I,b]||<\infty$. We used the fact that $\ran IQ\subset\ran
P(N\leq 1)$, since $I$ is linear in creators and $NQ=0$.\hfill $\blacksquare$\\

We define the operator $[L,A]$ by $\dom([L,A])=\dom(N)$ and
\begin{equation}
[L,A]:=[L,A_0]+[L,b]=N+\lambda\tilde{I}+[L,b].
\label{034}
\end{equation}
Again, we point out that $[L,A]$ is to be understood as the r.h.s. of \fer{034} (with $[L,b]$ defined in Proposition 4.2). The commutator notation $[L,A]$ is chosen because in the sense of quadratic forms on $\dom(L_0)\cap\dom(N)\cap\dom(A_0)$, one has $\scalprod{\varphi}{[L,A]\varphi}=2\RE\scalprod{L\varphi}{A\varphi}$ with $A=A_0+b$. Define now the operator $B$ by $\dom(B)=\dom(N)$ and
\begin{equation}
B:=[L,A]-\frac{1}{10}N=\frac{9}{10}N+\lambda\tilde{I}+[L,b].
\label{B}
\end{equation}
Here is the main result of this section:\\

{\bf Theorem 4.3\ }{\it Let $e\in\sigma(L_p)$ and let $\Delta$ be an interval around $e$
not containing any other eigenvalue of $L_p$. Let $E_\Delta$ be the (sharp) indicator
function of $\Delta$ and set  $\edeltanot=E_\Delta(L_0)$. Assume that the
Fermi Golden Rule Condition \fer{fgrc} (or \fer{fgrc0}) holds. Then there is a number $s>0$ s.t. if $0<\theta,\epsilon,\epsilon\theta^{-1},\theta\lambda^2\epsilon^{-3}<s$, then we have on $\dom(N^{1/2})$, in the sense of quadratic forms:
\begin{equation}
\edeltanot B\edeltanot\geq\frac{\theta\lambda^2}{\epsilon}\gamma_e\edeltanot
(1-{\textstyle {\frac{5}{2}}}\delta_{e,0}P_{\Omega_{\beta,0}})\edeltanot,
\label{3.4}
\end{equation}
where $P_{\Omega_{\beta,0}}$ is the projector onto the span of
$\Omega_{\beta,0}$ defined in \fer{XXp}.
}\\

An essential ingredient of the  proof of Theorem 4.3 is the {\it Feshbach method}, which we explain now.

\subsection{The Feshbach method}
The main idea of the Feshbach method is to use an isospectral correspondence between operators acting on a  Hilbert space and operators acting on some subspace.  We explain this method adapted
to our case. For a more general exposition, see e.g. [BFS2] and [DJ].\\
\indent
Consider the Hilbert spaces $\h_e$ defined by
$
 \h_e=\ran\chi_\nu\edeltanot,
$
where $\chi_\nu=\chi(N\leq\nu)$ is a cutoff in $N$, and $\nu$ is a positive integer. With our definitions of $Q, \qbar$, (see \fer{Q}) we have
\begin{equation}
\h_e=\ran\chi_\nu\edeltanot Q\oplus \ran\chi_\nu\edeltanot\qbar.
\label{cutoff}
\end{equation}
Define $Q_1=\chi_\nu\edeltanot Q$ and $Q_2=\chi_\nu\edeltanot\qbar$ and set $B_{ij}=Q_iBQ_j$,
$i,j=1,2$. The operators $B_{ij}$ are bounded due to the cutoff in $N$. Notice that $Q_{1,2}$ are projection operators (i.e. $Q_{1,2}^2=Q_{1,2}$) since $\chi_\nu$ commutes with $\edeltanot$ and $Q$.\\
\indent
The main ingredient of the Feshbach method is the following
observation:\\

{\bf Proposition 4.4 (isospectrality of the Feshbach map).\ }{\it If $z$ is in the
resolvent set of $B_{22}$ (i.e. if $(B_{22}-z)^{-1}\upharpoonright\ran Q_2$ exists as a bounded operator) and if
\begin{equation}
\norm{Q_2(B_{22}-z)^{-1}Q_2BQ_1}<\infty,\ \ \norm{Q_1BQ_2(B_{22}-z)^{-1}Q_2}<\infty,
\label{FA}
\end{equation}
then we have $z\in\sigma_{\#}(B) \Longleftrightarrow z\in\sigma_{\#}({\cal E}_z)$, where
the Feshbach map ${\cal E}_z={\cal E}_z(B)$ is defined by
$
B\mapsto {\cal E}_z=B_{11}-B_{12}(B_{22}-z)^{-1}B_{21},
$
and $\sigma_{\#}$ stands for $\sigma$ or $\sigma_{pp}$ (spectrum or pure point spectrum).}\\

The proof of Proposition 4.4 is given in a more general setting e.g. in [BFS2], [DJ],  we do not
repeat it here. We use the isospectrality of the Feshbach map to show positivity of $B$
in the following way (see also [BFSS]):
\\

{\bf Corollary 4.5\ }{\it Let $\vartheta_0=\inf\sigma(B\upharpoonright\h_e)$
and suppose that $B_{22}\geq\tilde{\vartheta}Q_2$ for some $\tilde{\vartheta}>-\infty$, and that $\inf\sigma({\cal
E}_\vartheta)\geq \Sigma_0$ uniformly in $\vartheta$ for $\vartheta\leq \vartheta_1$, where $\Sigma_0$ and $\vartheta_1$ are two fixed (finite) numbers. Then we have 
$\vartheta_0\geq\min\{\tilde{\vartheta},\inf\sigma({\cal E}_{\vartheta_0})\}$.
}
\\

{\it Remarks.\ } 1.\ All our estimates in this section will be independent of the $N$-cutoff introduced in \fer{cutoff}. In particular, $\tilde{\vartheta}, \vartheta_0, \vartheta_1, \Sigma_0$ are independent of $\nu$. This will allow us to obtain  inequality \fer{3.4} on $\dom(N^{1/2})$ from the corresponding estimate on $\ran \chi(N\leq\nu)$ by letting $\nu\rightarrow \infty$ (see \fer{limit} below).\\
\indent
2.\ The condition  $\inf\sigma({\cal E}_\vartheta)\geq \Sigma_0$ uniformly in $\vartheta$ for
$\vartheta\leq \vartheta_1$, implies  that $\vartheta_0\neq -\infty$.\\

{\it Proof of Corollary 4.5.\ } If $\vartheta_0>\tilde{\vartheta}$, then the assertion is clearly true. If $\vartheta_0<\tilde{\vartheta}$, then
$\vartheta_0$ is in the resolvent set of $B_{22}$, and it is easy to show that \fer{FA} holds for $z=\vartheta_0$, so $\vartheta_0\in\sigma({\cal E}_{\vartheta_0})$, i.e. $\vartheta_0\geq\inf\sigma({\cal E}_{\vartheta_0})$.\hfill $\blacksquare$

\subsection{Proof of Theorem 4.3 (using the Feshbach method).}

We apply Corollary 4.5 to the operator
\begin{equation}
B'=B-\delta_{e,0}\delta\pfinotperp,
\label{15.1}
\end{equation}
where $\delta_{e,0}$ is the Kronecker symbol, i.e. $\delta_{e,0}$ is one if $e=0$ and zero else. The positive number $\delta$ will be chosen appropriately below, see after \fer{499}.

 First, we show that $B'_{22}\geq (3/4-\delta_{e,0}\,\delta)Q_2$ (see \fer{B_22}), then we show that ${\cal E}_\vartheta\geq -1-\delta_{e,0}\,\delta=:\Sigma_0$ (see Proposition 4.6), uniformly in $\vartheta$ for $\vartheta \leq 1/2-\delta_{e,0}\,\delta$. Invoking Corollary 4.5 will then yield the result. Notice that due to the cutoff $\chi_\nu$ in \fer{cutoff}, $B_{ij}$, $i,j\in\{1,2\}$ are bounded operators. All the following estimates are independent of $\nu$.\\
\indent
We first calculate $B'_{22}=Q_2B'Q_2$. Using $QQ_2=0$, and $\delta_{e,0}\pfinotperp Q_2=\delta_{e,0}Q_2$, we obtain from \fer{15.1} and \fer{B}
\begin{equation}
B'_{22}=Q_2\left(\frac{9}{10}N +\lambda\tilde{I}+\theta\lambda^2(\repsilonbar^2IQI-IQI\repsilonbar^2)-\delta_{e,0}\delta\right)Q_2.
\label{b22}
\end{equation}
Proceeding as in the proof of Proposition A.1, one shows that $\forall c>0$,
\begin{equation*}
\left|\scalprod{\psi}{\lambda\tilde{I}\psi}\right|\leq c||N^{1/2}\psi||^2+C\textstyle{\frac{\lambda^2}{c}} ||\partial_ug_1||^2_{L^2}||\psi||^2.
\end{equation*}
With our assumptions on $g$, $||\partial_ug_1||^2_{L^2}<\infty$, uniformly in $\beta\geq 1$. Using the inequality above with $c=1/10$ and $\|\repsilonbar^2 IQI\|\leq C\epsilon^{-2}$, we obtain
\begin{equation*}
B'_{22}\geq Q_2\left( \frac{8}{10}N -O(\lambda^2 +\theta\lambda^2\epsilon^{-2})-\delta_{e,0}\delta\right) Q_2.
\end{equation*}
As can be easily checked, $Q_2=Q_2 \pomegaperp$, so we have $NQ_2\geq Q_2$, and we conclude that there is a $s_1>0$ s.t. if $\lambda^2+\theta\lambda^2\epsilon^{-2}\leq s_1$, then 
\begin{equation}
B'_{22}\geq\left(\frac{8}{10}-\delta_{e,0}\delta-O(\lambda^2+\theta\lambda^2\epsilon^{-2})\right)Q_2\geq\left(\frac{3}{4}-\delta_{e,0}\delta\right)Q_2.
\label{B_22}
\end{equation}
In the language of Corollary 4.5, this means we can take $\tilde{\vartheta}=3/4-\delta_{e,0}\delta$.\\
\indent
In a  next step, we calculate a lower bound on ${\cal E}_\vartheta$ for $\vartheta\leq 1/2-\delta_{e,0}\delta$.\\

\noindent
{\bf Proposition 4.6\ }{\it We have, uniformly in $\vartheta$ for $\vartheta\leq 1/2-\delta_{e,0}\delta$:
\begin{equation}
{\cal E}_\vartheta\geq 2\pi\frac{\theta\lambda^2}{\epsilon}(1-5\theta) Q_1\left( \Gamma(e)-\frac{\epsilon\delta_{e,0}\delta}{2\theta\lambda^2}\ppperp-O(\epsilon^{1/4}+\epsilon\theta^{-1}+\theta\lambda^2\epsilon^{-3})\right)Q_1,
\label{15}
\end{equation}
where the error term is independent of $\delta$. Recall that $\Omega^p_\beta$
is the particle Gibbs state defined in \fer{particlegibbs}.
}\\

{\it Proof of Proposition 4.6\ } By definition, ${\cal E}_\vartheta =B'_{11}-B'_{12}(B'_{22}-\vartheta)^{-1}B'_{21}$. We show that $B'_{11}$ is positive and $B'_{12}(B'_{22}-\vartheta)^{-1}B'_{21}$ is small compared to $B'_{11}$.\\
\indent
 With $\qbar Q_1=0$, $QQ_1=Q_1$ and $\delta_{e,0}\pfinotperp Q_1=\delta_{e,0}\ppperp Q_1$, we obtain from \fer{15.1} and \fer{B}:
\begin{equation}
B'_{11}\geq  2\theta\lambda^2 Q_1 \left(I\repsilonbar^2 I-\frac{\delta_{e,0}\delta}{2\theta\lambda^2}\ppperp\right)Q_1-O(\lambda^2),\label{5}
\end{equation}
where we used $\lambda\tilde{I}\geq-\frac{1}{10}N-O(\lambda^2)$ and $Q_1N=0$.\\
\indent
Let us now examine $B'_{12}(B'_{22}-\vartheta)^{-1}B'_{21}$. Notice that from \fer{b22}, we get
\begin{equation}
Q_2(B'_{22}-\vartheta)Q_2 =\textstyle{\frac{9}{10}}Q_2N^{1/2}(1-\textstyle{\frac{10}{9}}(\vartheta+\delta_{e,0}\delta) N^{-1}+K_1)N^{1/2}Q_2,
\label{6}
\end{equation}
where we defined the bounded selfadjoint operator $K_1$ acting on $\ran Q_2$ as
\begin{equation}
K_1=\frac{10}{9}N^{-1/2}\left(\lambda\tilde{I} +\theta\lambda^2(\repsilonbar^2 IQI-IQI\repsilonbar^2)\right)N^{-1/2}.
\label{7}
\end{equation}
Since $||Q_2N^{-1/2}||\leq 1$ and $||\tilde{I}(N+1)^{-1/2}||\leq C$, we get
$||K_1||\leq C(\lambda +\theta\lambda^2\epsilon^{-2})$. Now on $\ran
P_\Omega^\perp$, we have $N\geq 1$, so since we look at $\vartheta$ s.t. $\vartheta+\delta_{e,0}\delta\leq 1/2$, we obtain
\begin{equation}
1-\frac{10}{9}(\vartheta+\delta_{e,0}\delta) N^{-1}\geq1-\frac{10}{9}\frac{1}{2}=\frac{4}{9}.
\label{9}
\end{equation}
Therefore we can rewrite \fer{6} as
\begin{eqnarray}
Q_2(B'_{22}-\vartheta)Q_2&=&\textstyle{\frac{9}{10}}Q_2 N^{1/2}\big(1-\textstyle{\frac{10}{9}}(\vartheta+\delta_{e,0}\delta) N^{-1}\big)^{1/2}(\bbbone+K_2)\nonumber\\
& &\times \big(1-\textstyle{\frac{10}{9}}(\vartheta+\delta_{e,0}\delta) N^{-1}\big)^{1/2} N^{1/2}Q_2,
\label{8}
\end{eqnarray}
where
$
K_2=\big(1-\textstyle{\frac{10}{9}}(\vartheta+\delta_{e,0}\delta) N^{-1}\big)^{-1/2}K_1 \big(1-\textstyle{\frac{10}{9}}(\vartheta+\delta_{e,0}\delta) N^{-1}\big)^{-1/2},
$
and $||K_2||\leq\frac{9}{4}||K_1||=O(\lambda+\theta\lambda^2\epsilon^{-2})\lless 1$. We have thus from \fer{8}:
\begin{eqnarray}
Q_2(B'_{22}-\vartheta)^{-1}Q_2&=&\frac{10}{9} Q_2 N^{-1/2}\big(\bbbone -\textstyle{\frac{10}{9}}(\vartheta+\delta_{e,0}\delta) N^{-1}\big)^{-1/2}K^2\nonumber\\
&&\times\big(\bbbone -\textstyle{\frac{10}{9}}(\vartheta+\delta_{e,0}\delta) N^{-1}\big)^{-1/2}N^{-1/2}Q_2,
\label{510}
\end{eqnarray}
where $K=(\bbbone +K_2)^{-1/2}$ is bounded and selfadjoint with $||K||^2=||K^2||=||(\bbbone+K_2)^{-1}||\leq\frac{1}{1-||K_2||}<2$. We have therefore, from \fer{510} and \fer{9}, and uniformly in $\vartheta$ for $\vartheta\leq 1/2-\delta_{e,0}\delta$:
\begin{eqnarray}
\scalprod{\psi}{B'_{12}(B'_{22}-\vartheta)^{-1}B'_{21}\psi}&=&\frac{10}{9}||K\big(\bbbone -\textstyle{\frac{10}{9}}(\vartheta+\delta_{e,0}\delta) N^{-1}\big)^{-1/2}N^{-1/2}B'_{21}\psi||^2\nonumber\\
&\leq &2\frac{10}{9}\frac{9}{4}||N^{-1/2}B'_{21}\psi||^2 = 5||N^{-1/2}B_{21}\psi||^2.
\label{511}
\end{eqnarray}
Notice that $B'_{12}=B_{12}$ and $B'_{21}=B_{21}$.
Now, remembering \fer{B}, and since $NQ_1=0$ and $Q_2Q=0=\qbar Q_1$,
\begin{equation*}
N^{-1/2} B_{21}= N^{-1/2}Q_2\big[\lambda\tilde{I} +\theta\lambda(L_0-e)\repsilonbar^2I-\theta\lambda\repsilonbar^2I(L_0-e) +\theta\lambda^2(I\repsilonbar^2I-\repsilonbar^2IQI)\big]Q_1.
\end{equation*}
Using $||N^{-1/2}Q_2||\leq 1$, $||\tilde{I}Q_1||\leq C$, $||IQ_1||\leq C$,
$\|N^{-1/2}I\|\leq C$, $(L_0-e)Q_1=0$, $||(L_0-e)\repsilon||\leq 1$, we get
\begin{equation*}
||N^{-1/2}B_{21}\psi||^2\leq C(\lambda^2+\theta^2\lambda^4\epsilon^{-4})||\psi||^2+2\theta^2\lambda^2||\,\repsilonbar IQ_1\psi||^2,
\end{equation*}
thus with \fer{511}, we obtain
\begin{equation*}
-\scalprod{\psi}{B'_{12}(B'_{22}-\vartheta)^{-1}B'_{21}\psi}\geq -10\theta^2\lambda^2\scalprod{\psi}{Q_1I\repsilonbar^2 IQ_1\psi}-O(\lambda^2+\theta^2\lambda^4\epsilon^{-4})||\psi||^2,
\end{equation*}
and so, together with \fer{5}, we get, uniformly in $\vartheta$ for $\vartheta\leq 1/2-\delta_{e,0}\delta$:
\begin{equation}
{\cal E}_\vartheta \geq 2\theta\lambda^2(1-5\theta)Q_1(I\repsilonbar^2 I-\frac{\delta_{e,0}\delta}{2\theta\lambda^2}\ppperp)Q_1 -O(\lambda^2+\theta^2\lambda^4\epsilon^{-4}).
\label{10}
\end{equation}
We point out that the error term in the last inequality does not depend on
$\delta$. With the choice of parameters we will make (see \fer{parameters}),  \fer{10}  shows
that ${\cal E}_\vartheta\geq -1-\delta_{e,0}\delta$ uniformly in $\vartheta$
for $\vartheta\leq 1/2-\delta_{e,0}\delta$, i.e. in the language of Corollary 4.5, $\Sigma_0=-1-\delta_{e,0}\delta$.\\
\indent
The remaining part of the proof consists in relating the strict positivity of the nonnegative operator $Q_1 I\repsilonbar^2 IQ_1$ to the Fermi Golden Rule Condition. We let $I_a$ and $I_c=I^*_a$ denote the parts of $I$ containing annihilators and creators only, so that $I=I_a+I_c$. Thus
\begin{equation}
Q_1I\repsilonbar^2IQ_1=Q_1I_a\repsilonbar^2I_cQ_1=Q_1I_a\repsilon^2I_c Q_1.
\label{11}
\end{equation}
In the first step, we used $I_aQ_1=0$ and $Q_1I_c=0$ (since $I_aP_\Omega=0$)
and in the second step, we used $Q_1I_a\qbar=Q_1I_a$ (since $I_a Q=0$). Now write
\begin{equation}
Q_1I_a\repsilon^2 I_cQ_1=
Q_1\int\int m^*(u,\alpha)a(u,\alpha)\repsilon^2(e)a^*(u',\alpha')m(u',\alpha')\ Q_1,
\label{12}
\end{equation}
where $m$ is defined  \fer{m}, and where we display the dependence of $\repsilon^2$ on $e$. The operator-valued distributions ($a$ and $a^*$) satisfy the canonical commutation relations
$
[a(u,\alpha),a^*(u',\alpha')]=\delta(u-u')\delta(\alpha-\alpha').
$
Next, we notice that the pull-through formula
$
a(u,\alpha)L_f=(L_f+u)a(u,\alpha)
$
implies 
\begin{equation}
a(u,\alpha)\repsilon^2(e)=\repsilon^2(e-u)a(u,\alpha).
\label{019}
\end{equation}
Using the CCR and formula \fer{019} together with the fact that
$a(u,\alpha)Q_1=0$, we commute $a(u,\alpha)$ in \fer{12} to the right and arrive at
\begin{equation}
\fer{12}= Q_1\int m^*(u,\alpha)R_\epsilon^2(e-u) m(u,\alpha)\ Q_1.
\label{14}
\end{equation}
We can pull a factor $P_\Omega$ out of $Q_1$ and place it inside the integral next to $\repsilon^2(e-u)$ and thus replace $\repsilon^2(e-u)$ by $((L_p-e+u)^2+\epsilon^2)^{-1}$. Notice that $\epsilon((L_p-e+u)^2+\epsilon^2)^{-1}\rightarrow\delta(L_p-e+u)$ as $\epsilon\rightarrow 0$. More precisely, we have\\

\noindent
{\bf Proposition 4.7\ }{\it There is an $s_2>0$ s.t. for $0<\epsilon<s_2$, we have
\begin{equation*}
Q_1\int m^*(u,\alpha) \left((L_p-e+u)^2+\epsilon^2\right)^{-1}m(u,\alpha)\
Q_1\geq Q_1\frac{\pi}{\epsilon}\left(\Gamma(e)-O(\epsilon^{1/4})\right)Q_1.
\end{equation*}
}
Proposition 4.7, which we prove in Appendix A.3, together with \fer{10}-\fer{12} and \fer{14} yields \fer{15}, proving Proposition 4.6.\hfill $\blacksquare$\\

Now we finish the proof of Theorem 4.3. If the Fermi Golden Rule Condition \fer{fgrc} holds, then for $e\neq 0$, we have $\Gamma(e)\geq \gamma_e>0$ on $\ran Q_1$, so we obtain from \fer{15}, and under the conditions on the parameters stated in Theorem 4.3: ${\cal E}_\vartheta\geq\pi\frac{\theta\lambda^2}{\epsilon}\gamma_e$, so by Corollary 4.5:
\begin{equation}
\inf\sigma(B\upharpoonright \h_e)\geq\min\{1/2,\pi\theta\lambda^2\epsilon^{-1}\gamma_e\}=\pi\frac{\theta\lambda^2}{\epsilon}\gamma_e,
\label{500}
\end{equation}
since by our choice of the parameters (see \fer{parameters}), we will have $\frac{\theta\lambda^2}{\epsilon}<(2\pi\gamma_e)^{-1}$.\\
\indent
For $e=0$, we have $\Gamma(0)=\Gamma(0)\ppperp$, since
$\Gamma(0)\Omega^p_\beta=0$  (see Theorem 2.4), so Proposition 4.6 gives
\begin{equation}
{\cal E}_\vartheta\geq \pi\frac{\theta\lambda^2}{\epsilon}Q_1\left(\left\{\gamma_0-\frac{\epsilon\delta}{2\theta\lambda^2}\right\}\ppperp -O(\epsilon^{1/4}+\epsilon\theta^{-1}+\theta\lambda^2\epsilon^{-3})\right)Q_1.
\label{499}
\end{equation}
For some fixed $0<a<\frac{\gamma_0}{2(\pi-1)}$ (independent of $\theta, \lambda, \epsilon)$, there is a $s_3>0$ s.t. if $0<\theta\lambda^2\epsilon^{-1}<s_3$, then $\gamma_0-\frac{\epsilon\delta}{2\theta\lambda^2}>-a$, which gives with \fer{499}:
\begin{eqnarray*}
{\cal E}_\vartheta&\geq& \pi\frac{\theta\lambda^2}{\epsilon}Q_1\left(-a\ppperp- O(\epsilon^{1/4}+\epsilon\theta^{-1}+\theta\lambda^2\epsilon^{-3})\right) Q_1\\
&\geq&\pi\frac{\theta\lambda^2}{\epsilon}\left( -a-O(\epsilon^{1/4}+\epsilon\theta^{-1}+\theta\lambda^2\epsilon^{-3})\right)Q_1\\
&\geq&-2\pi\frac{\theta\lambda^2}{\epsilon}a\,Q_1.
\end{eqnarray*}
The last step is true provided $\epsilon^{1/4}+\epsilon\theta^{-1}+\theta\lambda^2\epsilon^{-3}<s_4$, for some small $s_4>0$. 
Remembering that $B'=B-\delta \pfinotperp$, we obtain from Corollary 4.5
\begin{equation*}
\inf\sigma\left((B-\delta\pfinotperp)\upharpoonright\h_0\right)\geq\min\{1/2,-2\pi a \theta\lambda^2/\epsilon\}=-2\pi a \frac{\theta\lambda^2}{\epsilon},
\end{equation*}
from which we conclude that if the condition on the parameters given in Theorem 4.3 is satisfied with $s=\min(s_1,s_2,s_3,s_4)$, then 
\begin{eqnarray}
\chi_\nu \edeltanot B\edeltanot\,\chi_\nu  &\geq& \chi_\nu \edeltanot\left(-2\pi a \frac{\theta\lambda^2}{\epsilon}+\delta\pfinotperp\right)\edeltanot\,\chi_\nu \nonumber\\
&=&2\frac{\theta\lambda^2}{\epsilon}\gamma_0\chi_\nu \edeltanot\left(1-a(\pi-1)/\gamma_0-(1+a/\gamma_0)P_{\Omega_{\beta,0}}\right)\edeltanot\,\chi_\nu \nonumber\\
&\geq&\frac{\theta\lambda^2}{\epsilon}\gamma_0\chi_\nu \edeltanot (1-{\textstyle{\frac{5}{2}}}P_{\Omega_{\beta,0}})\edeltanot\,\chi_\nu ,
\label{501}
\end{eqnarray}
where we used $a/\gamma_0\leq\frac{1}{2(\pi-1)}$. Estimates \fer{500} and \fer{501} yield $\forall \psi$:
\begin{equation}
\scalprod{\psi}{\chi_\nu\edeltanot B\edeltanot \chi_\nu\psi}\geq\frac{\theta\lambda^2}{\epsilon}\gamma_e\scalprod{\psi}{\chi_\nu\edeltanot(1-{\textstyle{\frac{5}{2}}}\delta_{e,0}P_{\Omega_{\beta,0}})\edeltanot\chi_\nu\psi}.
\label{limit}
\end{equation}
Suppose now $\psi\in\dom(N^{1/2})$. Then, since $(N+1)^{-1/2} B (N+1)^{-1/2}$ is bounded (see the definition of $B$, \fer{B}), and since $\chi_\nu\rightarrow \bbbone$ strongly as $\nu\rightarrow\infty$, we conclude that $\forall \psi\in\dom(N^{1/2})$:
\begin{equation*}
\scalprod{\psi}{\edeltanot B\edeltanot\psi}\geq\frac{\theta\lambda^2}{\epsilon}\gamma_e\scalprod{\psi}{\edeltanot(1-{\textstyle{\frac{5}{2}}}\delta_{e,0}P_{\Omega_{\beta,0}})\edeltanot\psi},
\end{equation*}
which proves Theorem 4.3.\hfill $\blacksquare$

\section{Proof of Theorem 2.1: step 2.}

We pass from the positive commutator estimate w.r.t. $L_0$ given in Theorem
4.3 to one w.r.t. the full Liouvillian $L$, hence proving Theorem 2.1. The essential ingredient of this
procedure is the IMS localization formula, which we apply to a partition of
unity w.r.t. $N$. Then, we carry out the estimates on each piece of the partition separately.

\subsection{PC with respect to spectral localization in $L$}

 Let
$
1=\hat{\chi}^2_1(x)+\hat{\chi}^2_2(x)$, $x\in{\mathbb R}_+$,
$\hat{\chi}_1^2\in C_0^\infty([0,1])$, be a $C^\infty$-partition of unity. For some  scaling parameter $\sigma\ggeq 1$, define
$
\chi_i=\chi_i(N)=\hat{\chi}_i(N/\sigma),\ \ \ i=1,2.
$
The reason why we introduce the partition of unity is that $I\chi_1=O(\sigma^{1/2})$ is bounded.
Since the $\chi_i$ leave $\dom(N^{1/2})$ invariant, then
$
[\chi_i,[\chi_i,B]]=\chi_i^2B-2\chi_iB\chi_i+B\chi_i^2
$
is well defined on $\dom(N^{1/2})$ in the sense of quadratic forms, and by summing over $i=1,2$, we get the so-called IMS localization formula (see also [CFKS]):
\begin{equation}
B=\sum_{1,2}\chi_iB\chi_i+\frac{1}{2}[\chi_i,[\chi_i,B]].
\label{IMS}
\end{equation}
Furthermore, we obtain from \fer{IMS} and \fer{B}, in the sense of quadratic
forms on $\dom(N^{1/2})$:
\begin{equation}
h(L)[L,A]h(L)=\frac{1}{10}h(L)Nh(L)+\sum_{1,2}h(L)\chi_iB\chi_ih(L)+\frac{1}{2}h(L)[\chi_i,[\chi_i,B]]h(L).
\label{28}
\end{equation}
In Propositions 5.1-5.3 below, we estimate the different terms on the r.h.s. of \fer{28}. Then we complete the proof of Theorem 2.1 by choosing suitable relations among the parameters $\theta, \lambda, \epsilon, \sigma$ (see \fer{parameters}).\\

\indent
{\bf Proposition 5.1.\ }{\it There is a $s_5>0$ s.t. if $\lambda^2\sigma^{-1}<s_5$, then
\begin{equation}
h\chi_2B\chi_2 h \geq\frac{\sigma}{2}h\chi_2^2h.
\label{29}
\end{equation}}

\noindent
{\it Proof.\ } Recall that $B=\frac{9}{10} N+\lambda\tilde{I}+[L,b]$. Since $Q\chi_2=0$ and $QI\chi_2=0$ (see also end of proof of Proposition 4.2), we have $\forall\psi$: $\scalprod{\psi}{\chi_2[L,b]\chi_2\psi}=0$. 
Furthermore, Proposition 6.1 gives $\forall c>0$, $\lambda\tilde{I}\geq cN-O(\lambda^2/c)$, so 
\begin{equation*}
\scalprod{\psi}{\chi_2(9N/10+\lambda\tilde{I})\chi_2\psi}\geq\scalprod{\psi}{\chi_2\left[(9/10-c)N-O(\lambda^2/c)\right]\chi_2\psi}\geq\displaystyle{\frac{3}{4}}\sigma\scalprod{\psi}{\chi_2^2\psi},
\end{equation*}
provided $\lambda^2\sigma<s_5$ and 
where we picked the value $c=1/10$ and used $\chi_2N\chi_2\geq\sigma\chi_2^2$.\hfill $\blacksquare$\\

{\bf Proposition 5.2.\ }{\it We have
\begin{eqnarray*}
h\chi_1B\chi_1h +\frac{1}{10}h Nh
&\geq& \frac{\theta\lambda^2}{\epsilon}\gamma_e\left(1-O(\lambda\sigma^{1/2})\right)h\chi_1^2h -\frac{5}{2}\frac{\theta\lambda^2}{\epsilon}\gamma_0\delta_{e,0} h P_{\Omega_{\beta,0}} h \\
&&-\frac{\theta\lambda^2}{\epsilon}O\left(\epsilon\theta^{-1}+\epsilon\sigma^{1/2}+\lambda\sigma\epsilon^{-1}\right)h^2.
\end{eqnarray*}}

{\it Proof.\ } Let $\edprime:=F_{\Delta'}(L_0)$, where $\Delta'$ is an interval whose interior contains the closure of $\Delta$, and $F_{\Delta'}$ is a smooth characteristic function with support in $\Delta'$, s.t. $E_\Delta(L_0)\overline{\edprime}=0$, where we denoted $\bbbone-\edprime=:\overline{\edprime}$. We take $\Delta'$ to contain only one eigenvalue of $\sigma(L_0)$, namely $e$, so that \fer{3.4} in  Theorem 4.3  holds, with $E_\Delta^0$ replaced by $E_{\Delta'}^0$. We have
\begin{eqnarray}
h \chi_1B\chi_1 h+\frac{1}{10}h N h&=&h\chi_1\edprime B\edprime\chi_1 h \label{30}\\
& &+\frac{1}{20}h N h+h\chi_1\edprime B\overline{\edprime}\chi_1 h +\mbox{\ adjoint} \label{31}\\
& &+h\chi_1\overline{\edprime} B\overline{\edprime}\chi_1 h.
\label{32}
\end{eqnarray}
First, we show that \fer{31} and \fer{32} are bounded below by small terms. To treat \fer{31}, notice that
\begin{eqnarray}
\chi_1\edprime B\overline{\edprime}\chi_1&=&\chi_1\edprime(9N/10+\lambda\tilde{I}+[L,b])\overline{\edprime}\chi_1\nonumber\\
&=&\frac{9}{10}\chi_1^2\edprime\overline{\edprime}N+\chi_1\edprime(\lambda\tilde{I}+[L,b])\overline{\edprime}\chi_1\nonumber\\
&\geq&\chi_1\edprime(\lambda\tilde{I}+[L,b])\overline{\edprime}\chi_1.
\label{020}
\end{eqnarray}
Now for $\phi_{1,2}\in\dom(N^{1/2})$, we have for any $c>0$ (see Proposition A.1)
\begin{eqnarray*}
\left|\scalprod{\phi_1}{\lambda\tilde{I}\phi_2}\right|&\leq&\lambda\left( \left|\scalprod{\phi_1}{\tilde{I}_a\phi_2}\right|+\left|\scalprod{\phi_2}{\tilde{I}_a\phi_1}\right|\right)\\
&\leq&C\lambda\left(||\phi_1||\,||N^{1/2}\phi_2|| +||\phi_2||\,||N^{1/2}\phi_1||\right)\\
&\leq&C\lambda^2c^{-1}\left(||\phi_1||^2+||\phi_2||^2\right)+ c\left(||N^{1/2}\phi_1||^2+||N^{1/2}\phi_2||^2\right).
\end{eqnarray*}
With $\phi_1=\edprime\chi_1\psi$, $\phi_2=\overline{\edprime}\chi_1\psi$, this yields $\forall c>0$:
\begin{equation*}
\left|\scalprod{\psi}{\chi_1\edprime\lambda\tilde{I}\overline{\edprime}\chi_1\psi}\right|\leq C\frac{\lambda^2}{c}2||\chi_1\psi||^2+2c||N^{1/2}\chi_1\psi||^2,
\end{equation*}
so
$
\chi_1\edprime\lambda\tilde{I}\overline{\edprime}\chi_1+\mbox{adjoint}\geq -4\left( C\frac{\lambda^2}{c}\chi_1^2+cN\right)$.
Taking $c<1/40$ gives then
\begin{eqnarray}
\frac{1}{20}h Nh +h\chi_1\edprime\lambda\tilde{I}\overline{\edprime}\chi_1h+\mbox{adjoint}&\geq&(1/10-4c)h Nh -C\lambda^2h\chi_1^2h\nonumber\\
&\geq& -C\lambda^2h\chi_1^2h.
\label{33}
\end{eqnarray}
Next, using $Q\overline{\edprime}=0$ and $(L_0-e)Q=0$, we calculate
\begin{eqnarray}
\chi_1\edprime[L,b]\overline{\edprime}\chi_1&=&\chi_1\edprime[L_0-e,b]\overline{\edprime}\chi_1 +\lambda\chi_1\edprime [I,b]\overline{\edprime}\chi_1\nonumber\\
&=&\theta\lambda\chi_1\edprime QI\repsilonbar^2(L_0-e)\overline{\edprime}\chi_1\nonumber\\
& &+\theta\lambda^2\chi_1\edprime\left(-\repsilonbar^2IQI -IQI\repsilonbar^2+QI\repsilonbar^2I\right)\overline{\edprime}\chi_1\nonumber\\
&=&O(\theta\lambda+\theta\lambda^2\epsilon^{-2}\sigma^{1/2}),
\label{34}
\end{eqnarray}
where we used $||\repsilon\overline{\edprime}||\leq|\Delta'|^{-1}\leq C$ and
$||I\chi_1||\leq C\sigma^{1/2}$. Next, since $\supp
h\cap\supp\overline{\edprime}=\emptyset$, then $\chi_1\overline{\edprime}
h(L)=\chi_1\overline{\edprime}(h(L)-h(L_0))$, so by using the operator
calculus introduced in Appendix A.4, we obtain
\begin{equation}
\chi_1\overline{\edprime}h(L)=\chi_1\int d\tilde{F}_{\Delta'}(z) (L_0-z)^{-1}\lambda I(L-z)^{-1}h(L) =O(\lambda\sigma^{1/2}).
\label{35}
\end{equation}
From \fer{34}, we then have
$
h\chi_1\edprime[L,b]\overline{\edprime}\chi_1h\geq-C\frac{\theta\lambda^2}{\epsilon}(\epsilon\sigma^{1/2}+\lambda\sigma\epsilon^{-1})h^2,
$
which, together with \fer{33} and \fer{020} yields
\begin{equation}
\fer{31}\geq -C\frac{\theta\lambda^2}{\epsilon}(\epsilon\theta^{-1}+\epsilon\sigma^{1/2}+\lambda\sigma\epsilon^{-1})h^2.
\label{36}
\end{equation}
Our next step is estimating \fer{32}. Again, using $Q\overline{\edprime}=0$, we get
\begin{eqnarray*}
\chi_1\overline{\edprime}B\overline{\edprime}\chi_1&=&\chi_1\overline{\edprime}(9N/10+\lambda\tilde{I})\overline{\edprime}\chi_1 -\theta\lambda^2\chi_1\overline{\edprime}\left(\repsilonbar^2IQI +IQI\repsilonbar^2\right)\overline{\edprime}\chi_1\\
&\geq&-C(\lambda^2+\theta\lambda^2),
\end{eqnarray*}
where we used $\lambda\tilde{I}\geq -cN-O(\lambda^2/c)$ and
$||\overline{\edprime}\repsilon^2||\leq |\Delta'|^{-2}\leq C$. We thus obtain,
since $\theta\lless 1$:
\begin{equation}
\fer{32}=h\chi_1\overline{\edprime}B\overline{\edprime}\chi_1h\geq -C\frac{\theta\lambda^2}{\epsilon}\frac{\epsilon}{\theta}h^2.
\label{37}
\end{equation}
Finally, we investigate the positive term \fer{30}. By sandwiching \fer{3.4} in  Theorem 4.3 (with $E_\Delta^0$ replaced by $E_{\Delta'}^0$) with $\edprime$, and noticing that $\edprime E_{\Delta'}^0=\edprime$, we arrive at
\begin{eqnarray}
h\chi_1\edprime B\edprime\chi_1h&\geq&\pi\frac{\theta\lambda^2}{\epsilon}\gamma_eh \chi_1\edprime \left(1-{\textstyle{\frac{5}{2}}}\delta_{e,0}P_{\Omega_{\beta,0}}\right)\edprime\chi_1 h \nonumber\\
&\geq&\frac{\theta\lambda^2}{\epsilon}\gamma_e h \left(\chi_1^2(\edprime)^2 -{\textstyle{\frac{5}{2}}}\delta_{e,0} P_{\Omega_{\beta,0}}\right) h   \nonumber\\
&=&\frac{\theta\lambda^2}{\epsilon}\gamma_eh\left(\chi_1^2\left(1-\overline{\edprime}\right)^2-{\textstyle{\frac{5}{2}}}\delta_{e,0} P_{\Omega_{\beta,0}}\right) h\nonumber\\
&\geq& \frac{\theta\lambda^2}{\epsilon}\gamma_eh\left(\chi_1^2\left(1-2\overline{\edprime}\right) -{\textstyle{\frac{5}{2}}}\delta_{e,0} P_{\Omega_{\beta,0}}\right) h\nonumber\\
&\geq& \frac{\theta\lambda^2}{\epsilon}\gamma_eh\left( \chi_1^2(1-C\lambda\sigma^{1/2})-{\textstyle{\frac{5}{2}}}\delta_{e,0} P_{\Omega_{\beta,0}}\right)h,
\label{513}
\end{eqnarray}
where we used \fer{35} in the last step once again, and  $-2\chi_1^2(\edprime)^2P_{\Omega_{\beta,0}}\geq -2P_{\Omega_{\beta,0}}$ in the second step. Combining \fer{513} with \fer{36} and \fer{37} yields Proposition 5.2.\hfill  $\blacksquare$\\

{\bf Proposition 5.3.\ }{\it We have
$
\sum_{1,2}h[\chi_i,[\chi_i,B]]h=\frac{\theta\lambda^2}{\epsilon}O (\epsilon\theta^{-1}\lambda^{-1}\sigma^{-3/2} )h^2.
$}\\

{\it Proof.\ } Notice that $\chi_1$ and $1-\chi_2$ have compact supports contained in $[0,2]$. Now in the double commutator, we can replace $\chi_2$ by $1-\chi_2$ without changing its value. So if suffices to estimate $[\chi(N/\sigma),[\chi(N/\sigma),B]]$, where $\chi\in C_0^\infty([0,2])$. We have
$
[\chi(N/\sigma),[\chi(N/\sigma),B]]=[\chi(N/\sigma),[\chi(N/\sigma),\lambda\tilde{I}+[L,b]]]$. It is not difficult to see that 
we have in the sense of operators on $\dom(N^{1/2})$:
\begin{eqnarray}
[\chi(N/\sigma),[\chi(N/\sigma),\lambda\tilde{I}]]&=&\frac{\lambda}{\sigma^2}\int d\tilde{\chi}(z)\int d\tilde{\chi}(\zeta)(N/\sigma-z)^{-1}(N/\sigma-\zeta)^{-1}\nonumber\\
& &\ \ \ \ \times\tilde{I}(N/\sigma-z)^{-1}(N/\sigma-\zeta)^{-1}.
\label{38}
\end{eqnarray}
We used the operator calculus introduced in Appendix A.4. 
Now since $\|\tilde{I}(N/\sigma-z)^{-1/2}\|\leq C \|(N+1)^{1/2}(N/\sigma-z)^{-1}\|\leq C\sigma^{1/2}|\IM z|^{-1}$, which follows from
\begin{equation*}
\sup_{x\geq 0}\frac{\sqrt{x+1}}{|x/\sigma-z|}\leq C\sigma^{1/2}|\mbox{Im}z|^{-1},
\end{equation*}
we conclude that
\begin{equation}
\left\|[\chi(N/\sigma),[\chi(N/\sigma),\lambda\tilde{I}]]\right\|\leq C\lambda\sigma^{-3/2}.
\label{39}
\end{equation}

Next, write for simplicity $\chi$ instead of $\chi(N/\sigma)$, and look at 
\begin{equation*}
[\chi,[\chi,[L,b]]]=\theta\lambda[\chi,[\chi,[L,\repsilonbar^2IQ]]]+\mbox{adjoint}.
\end{equation*}
 We claim that
\begin{equation}
[\chi,[L,\repsilonbar^2IQ]]=0.
\label{=0}
\end{equation}
Write first $[L,\repsilonbar^2IQ]=\repsilonbar^2[L_0,I]Q+\lambda [I,\repsilonbar^2IQ]$. Then  $[\chi,\repsilonbar^2[L_0,I]Q]=[\overline{\chi},\repsilonbar^2[L_0,I]Q]=\overline{\chi}\repsilonbar^2[L_0,I]Q-\repsilonbar^2[L_0,I]Q\overline{\chi}$. Here, $\overline{\chi}=1-\chi$. Notice that $Q\overline{\chi}=0$, and since $\ran\repsilonbar^2[L_0,I]Q\subset\ran P(N=1)$, we have also $\overline{\chi}\repsilonbar^2[L_0,I]Q=0$, for $\sigma>2$. Similarly, $[\chi,[I,\repsilonbar^2IQ]]=0$, so \fer{=0} follows.\\
\indent
We obtain thus from \fer{39}: $[\chi,[\chi,B]]=O(\lambda\sigma^{-3/2})$, which proves the proposition.\hfill $\blacksquare$\\
\indent
Now we finish the proof of Theorem 2.1. The IMS localization formula \fer{28} together with Propositions 5.1, 5.2, 5.3 yields
\begin{eqnarray*}
h[L,A]h&\geq& \frac{\theta\lambda^2}{\epsilon}\gamma_e\left(1-O(\lambda\sigma^{1/2})\right) h\chi_1^2h +\frac{\sigma}{2}h\chi_2^2h-\frac{5}{2}\frac{\theta\lambda^2}{\epsilon}\gamma_0\delta_{e,0}h P_{\Omega_{\beta,0}} h\\
&&-\frac{\theta\lambda^2}{\epsilon}O\left( \epsilon\theta^{-1}+\epsilon\sigma^{1/2}+\lambda\sigma\epsilon^{-1}+\epsilon\theta^{-1}\lambda^{-1}\sigma^{-3/2}\right) h^2.
\end{eqnarray*}
The sum of the first two terms on the r.h.s. is bounded below by
$
\frac{\theta\lambda^2}{\epsilon}\gamma_e\left( 1-O(\lambda\sigma^{1/2})\right) h^2,
$
so we get
\begin{eqnarray}
h[L,A]h&\geq&\frac{\theta\lambda^2}{\epsilon} h \Big[ \gamma_e\left(1-{\textstyle{\frac{5}{2}}}\delta_{e,0}P_{\Omega_{\beta,0}}- O(\lambda\sigma^{1/2})\right)\nonumber\\
&&-O\left( \epsilon\theta^{-1}+\epsilon\sigma^{1/2}+\lambda\sigma\epsilon^{-1}+\epsilon\theta^{-1}\lambda^{-1}\sigma^{-3/2}\right)\Big] h.
\label{1.1}
\end{eqnarray}
Finally, we choose our parameters. Let $\epsilon=\lambda^{\hat{\epsilon}/100}$, $\sigma=\lambda^{-\hat{\sigma}/100}$, $\theta=\lambda^{\hat{\theta}/100}$, and choose
\begin{equation}
(\hat{\epsilon},\hat{\sigma},\hat{\theta})=(44,55,26).
\label{parameters}
\end{equation}
It is then easily verified that for small $\lambda$, the conditions on the parameters given in Theorem 4.3 and Proposition 5.1 hold, and furthermore, \fer{1.1} becomes
\begin{eqnarray*}
h[L,A]h&\geq&  \lambda^{182/100} h \left[\gamma_e\left( 1-{\textstyle{\frac{5}{2}}}\delta_{e,0}P_{\Omega_{\beta,0}}-O(\lambda^{145/200})\right)-O(\lambda^{1/200})\right] h\\
&\geq& \lambda^{91/50} h \left( \frac{\gamma_e}{2}(1-5\delta_{e,0}P_{\Omega_{\beta,0}}) -O(\lambda^{1/200})\right) h.\ \ \ \blacksquare
\end{eqnarray*}

\section{Proof of Theorem 2.2}

We follow the idea of the Virial Theorem, as explained in Subsection 1.3: 
Assume $\psi$ is a normalized eigenvector of $L$ with eigenvalue $e$. If $e=0$, we assume in addition that $\psi\in\ran P^\perp_{\Omega_{\beta,\lambda}}$. Let $\alpha>0$ and set 
$
f_\alpha:=\alpha^{-1}f(i\alpha A_0),
$
where $f$ is a bounded $C^\infty$-function,  such that the derivative $f'$ is
positive and s.t. $f'(0)=1$ (take e.g. $f=\Arctan$). Set 
\begin{equation*}
f'_\alpha:=f'(i\alpha A_0), \mbox{\ and \ } h_\alpha:=\sqrt{f'_\alpha}.
\end{equation*}
Furthermore, set $f''_\alpha:=f''(i\alpha A_0)$. For $\nu>0$ and $g\in C_0^\infty(-1,1)$, define $\psi_\nu=g(\nu N)\psi$. Here, $\alpha, \nu$ will be chosen small in an appropriate way. We define the
regularized eigenfunction $\psi_{\alpha,\nu}=h_\alpha\psi_\nu$. Notice that 
\begin{equation}
\psi_{\alpha, \nu}\rightarrow\psi, \ \ \mbox{as $\alpha, \nu\rightarrow 0$}. 
\label{l1}
\end{equation}
Set for notational convenience in this section
\begin{equation*}
K:=[L,A_0]=N+\lambda\tilde{I}.
\end{equation*}
The strategy is to show that $\av{K}_{\psi_{\alpha, \nu}}:=\scalprod{\psi_{\alpha,\nu}}{K\psi_{\alpha,\nu}}\rightarrow 0$, as $\alpha,\nu\rightarrow 0$ (see next Subsection, \fer{l5}). For this estimate, we need the restrictive IR behaviour $p>2$, see after Proposition 6.1. Using the PC estimate, Theorem 2.1, we also show that $\av{K}_{\psi_{\alpha,\nu}}$ is strictly positive (as $\alpha, \nu\rightarrow 0$), see Subsection 6.2, \fer{l15}. The combination of these two estimates yields a contradiction, hence showing that the eigenfunction $\psi$ of $L$ we started off with cannot exist.\\
\indent
In the case $e=0$, we need to use that the product 
$P_{\Omega_{\beta,0}}P^\perp_{\Omega_{\beta,\lambda}}$ is small, which is
satisfied provided $\beta|\lambda|<C$, see \fer{pertkms}.

\subsection{Upper bound on $\av{K}_{\psi_{\alpha,\nu}}$}

 Using $(L-e)\psi=0$ and that $[N,I]$ is $N^{1/2}$-bounded, we find that
\begin{equation}
 \av{f_\alpha(L-e)}_{\psi_\nu} = \av{g_\nu f_\alpha (L-e)g_\nu}_\psi=\av{f_\alpha g_\nu [\lambda I,g_\nu]}_\psi
=O(\lambda\alpha^{-1}\nu^{1/2}).
\label{l2}
\end{equation}
Next, observe that
\begin{equation}
2\IM\av{f_\alpha(L-e)}_{\psi_\nu}=\av{[L,if_\alpha]}_{\psi_\nu}
=\RE\av{[L,if_\alpha]}_{\psi_\nu}
=\RE\av{f'_\alpha N+\lambda[I,if_\alpha]}_{\psi_\nu},
\label{l3}
\end{equation}
where we used in the last step 
\begin{eqnarray*}
[L_0,if_\alpha]&=& \int d\tilde{f}(z)(i\alpha A_0-z)^{-1} [L_0,A_0] (i\alpha A_0-z)^{-1}\\
&=&\int d\tilde{f}(z) (i\alpha A_0-z)^{-2}N\\
&=&f'_\alpha N,
\end{eqnarray*}
since  $A_0$ and $N$ commute (second step) and we made use of \fer{deriv} with $p=1$ in the last step. The commutator $[I,if_\alpha]$ is examined in \\

{\bf Proposition 6.1.\ }{\it The following equality holds in the sense of operators on $\dom(N^{1/2})$ or in the sense of quadratic forms on $\dom(N^{1/4})$:
\begin{equation}
[I,if_\alpha]=f'_\alpha ad_{A_0}^1(I)-\frac{i}{2}\alpha f''_\alpha ad_{A_0}^2(I) +R,
\label{77.1}
\end{equation}
where we assume that the $k$-fold commutator
$
ad_{A_0}^k(I):=[\cdots[I,A_0],A_0,\cdots,A_0]
$
is $N^{1/2}$-bounded (or $N^{1/4}$-form bounded) for $k=1,2,3$. The term $R$ satisfies the estimate
$
RN^{-1/2},\ N^{-1/4}RN^{-1/4} =O(\alpha^2).
$
}\\

\noindent
{\it Proof.\ } Using the operator calculus introduced in Appendix A.4, we write 
\begin{eqnarray*}
[I,if_\alpha]&=&\int d\tilde{f}(z) (i\alpha A_0-z)^{-1} [I,A_0] (i\alpha A_0-z)^{-1}\\
&=& f'_\alpha ad_{A_0}^1(I)-i\alpha\int d\tilde{f}(z) (i\alpha A_0-z)^{-2} ad_{A_0}^2(I) (i\alpha A_0-z)^{-1}\\
&=& f'_\alpha ad_{A_0}^1(I) -\frac{i}{2}\alpha f''_\alpha ad_{A_0}^2(I)-\alpha^2\int d\tilde{f}(z) (i\alpha A_0-z)^{-3} ad_{A_0}^3(I) (i\alpha A_0-z)^{-1}.
\end{eqnarray*}
The last integral is defined to be $R$, and the estimates follow by noticing that $A_0$ and $N$ commute.\hfill $\blacksquare$\\
\indent
Notice that it is here that we need  $\| ad_{A_0}^k(I)N^{1/2}\|\leq C$, $k=2,3$, {\it hence the more restrictive IR behaviour $p>2$}. We obtain from \fer{77.1} and recalling that $\tilde{I}=[I,A_0]$:
\begin{eqnarray}
\fer{l3}&=&\RE\av{f'_\alpha K }_{\psi_\nu} -\frac{\lambda}{2}\RE\av{i\alpha f''_\alpha
  ad_{A_0}^2(I)}_{\psi_\nu} +O(\lambda\alpha^2\nu^{-1/2})\nonumber\\
&=&
\av{K}_{\psi_{\alpha,\nu}}+\lambda\RE\av{h_\alpha[h_\alpha,\lambda\tilde{I}]-\frac{i}{2}\alpha
  f''_\alpha ad_{A_0}^2(I)}_{\psi_\nu}
+O(\lambda\alpha^2\nu^{-1/2})\nonumber\\
&=&\av{K}_{\psi_{\alpha,\nu}}+O(\lambda\alpha^2\nu^{-1/2}).
\label{l4}
\end{eqnarray}
We used in the last step that the real part in the second term above is 
\begin{equation*}
\av{[h_\alpha,[h_\alpha,\tilde{I}]]-\frac{i}{2}\alpha [f''_\alpha,ad_{A_0}^2(I)]}_{\psi_\nu}=O(\alpha^2\nu^{-1/2}),
\end{equation*}
since $ad_{A_0}^3(I)$ is $N^{1/2}$-bounded.
 Combining  \fer{l4} and \fer{l2}, we obtain
\begin{equation}
\av{K}_{\psi_{\alpha,\nu}}\leq
C\lambda\left(\frac{\nu^{1/2}}{\alpha}+\frac{\alpha^2}{\nu^{1/2}}\right)\|\psi\|^2.
\label{l5}
\end{equation}

\subsection{Lower bound on $\av{K}_{\psi_{\alpha,\nu}}$}

 Let $\Delta$ be an interval containing exactly
one eigenvalue, $e$, of $L_p$. 
We introduce two partitions of unity. The first one is given by
\begin{equation*}
\chi_\Delta^2+\overline{\chi}_\Delta^2=1,
\end{equation*}
where $\chi_\Delta\in C^\infty(\Delta)$,
$\chi_\Delta(e)=1$. We localize in $L$, i.e. we set
$\chi_\Delta=\chi_\Delta(L)$.
The second partition of unity is given by 
\begin{equation*}
\chi^2+\overline{\chi}^2=1,
\end{equation*}
where $\chi\in C^\infty$ is a ``smooth Heaviside function'', i.e. $\chi(x)=0$ if
$x\leq 0$ and $\chi(x)=1$ if $x\geq 1$. We set for $n>0$: $\chi_n=\chi(N/n)$,
$\overline{\chi}_n^2=1- \chi_n^2$. We will choose $n<1/\nu$, so that
$
\chi_n\psi_\nu=\chi_n\psi$. 
The last equation will be used freely in what follows. We are going to use the IMS localization formula \fer{IMS} with respect to
both partitions of unity, and we start with the one localizing in $N$:
\begin{eqnarray}
\av{K}_{\psi_{\alpha,\nu}}&=&\av{\chi_nK\chi_n
  +\overline{\chi}_nK\overline{\chi}_n+\frac{1}{2}[\chi_n,[\chi_n,K]]+\frac{1}{2}[\overline{\chi}_n,[\overline{\chi}_n,K]]}_{\psi_{\alpha,\nu}}\nonumber\\
&\geq&\av{K}_{\chi_n \psi_{\alpha, \nu}}+\frac{n}{2}\|\overline{\chi}_n\psi_{\alpha, \nu}\|^2-O(\lambda n^{-3/2}),
\label{l6}
\end{eqnarray}
where we used that $K\geq n/2$ on $\ran P^\perp_\Omega$, and the estimate
\fer{39} with $\sigma$ replaced by $n$. 
Next, from the IMS localization
formula for the partition of unity w.r.t. $L$, we have
\begin{eqnarray}
\av{K}_{\chi_n\psi_{\alpha,\nu}}
&=&\av{\chi_\Delta
  K\chi_\Delta+\overline{\chi}_\Delta K\overline{\chi}_\Delta
  +R}_{\chi_n \psi_{\alpha,\nu}}\nonumber\\
&\geq&\av{\chi_\Delta(K+[L,b])\chi_\Delta +\overline{\chi}_\Delta K
  \overline{\chi}_\Delta +R}_{\chi_n\psi_{\alpha,\nu}}-\lambda^{19/50}O(\alpha
  n+\lambda n^{-1/2})\nonumber\\
&\geq& \theta\|\chi_\Delta\chi_n \psi_{\alpha,\nu}\|^2
-C\theta\delta_{e,0}\|P_{\Omega_{\beta,0}}\chi_\Delta\chi_n
\psi_{\alpha, \nu}\|^2 +\av{\overline{\chi}_\Delta K
  \overline{\chi}_\Delta +R}_{\chi_n \psi_{\alpha,\nu}}\nonumber\\
&&-\lambda^{19/50}O(\alpha n+\lambda n^{-1/2}).
\label{l7}
\end{eqnarray}
Here, several remarks are in order. First, we have set
$
2R=[\overline{\chi}_\Delta,[\overline{\chi}_\Delta,K]]+[\chi_\Delta,[\chi_\Delta,K]],
$
and we have used in the second step the fact that 
\begin{eqnarray*}
\av{[L,b]}_{\chi_\Delta\chi_n \psi_{\alpha,\nu}}&=&\av{[L-e,b]}_{\chi_\Delta\chi_n h_\alpha \psi} =2\RE\scalprod{\chi_\Delta(L-e)h_\alpha\chi_n\psi}{b\chi_\Delta\chi_n
  h_\alpha\psi}\\
&=&\lambda^{19/50}O(\alpha n+\lambda n^{-1/2}).
\end{eqnarray*}
We recall that $b$ is a bounded operator (see Proposition 4.2), with $\|b\|=O(\lambda^{19/50})$. 
In the last step in \fer{l7}, we used the positive commutator estimate,
Theorem 2.1, in the following way. For $e\neq 0$,  Theorem 4.1. gives right
away 
$
\chi_\Delta (K+[L,b])\chi_\Delta\geq\theta\chi^2_\Delta,
$
where we recall that $[L,A]=[L,A_0]+[L,b]$, and $b$ is defined in \fer{41.1}. We
have set $\theta=C\lambda^2$. In the zero eigenvalue case, $e=0$, we have 
\begin{eqnarray*}
\av{K+[L,b]}_{\chi_\Delta\chi_n\psi_{\alpha,\nu}}&\geq&\frac{\lambda^{91/50}}{2}\av{\gamma_0(1-5P_{\Omega_{\beta,0}})-O(\lambda^{1/200})}_{\chi_\Delta\chi_n\psi_{\alpha,\nu}}\\
&\geq&\frac{\lambda^{91/50}}{4}\gamma_0\|\chi_\Delta\chi_n \psi_{\alpha,\nu}\|^2 -\frac{5\lambda^{91/50}}{2}\gamma_0\|P_{\Omega_{\beta,0}}\chi_\Delta\chi_n \psi_{\alpha,\nu}\|^2.
\end{eqnarray*}
Setting again $\theta=C\lambda^{91/50}$ yields \fer{l7}.\\
\indent
We now estimate the remainder term $R$. Notice that the same observation as at
the beginning of the proof of Proposition 5.3 shows that we have the estimate
$
\av{R}_{\chi_n \psi_{\alpha,\nu}}=2i\IM\scalprod{\overline{\chi}_\Delta\chi_n\psi_{\alpha,\nu}}{[\overline{\chi}_\Delta,K]\chi_n\psi_{\alpha,\nu}}.
$
Therefore, 
\begin{equation}
\left|\av{R}_{\chi_n\psi_{\alpha,\nu}}\right|\leq
C\|\overline{\chi}_\Delta\chi_nh_\alpha\psi\|\,\|[\overline{\chi}_\Delta,
K]\chi_nh_\alpha\psi\|.
\label{l8}
\end{equation}
Now we have on $\dom(N)$:
$
[\overline{\chi}_\Delta, K]=\int d\tilde{\chi}_\Delta(z)
(L-z)^{-1}[K,L](L-z)^{-1},
$
where we recall that $(L-z)^{-1}$ leaves $\dom(N)$ invariant. Furthermore, 
\begin{equation}
[K,L]=\lambda[N,I]+\lambda[\tilde{I},L_0]+\lambda^2[\tilde{I},I]=\lambda [N,I]+\lambda I(u\partial_u g)+\lambda^2 [\tilde{I},I],
\label{l9}
\end{equation}
where $I(u\partial_u g)$ is obtained
from $I$ by replacing the form factor $g$ by $u\partial_ug$. The last
commutator in \fer{l9} is bounded, and the other two are $N^{1/2}$-bounded, so
we obtain
\begin{equation}
\left\|[\overline{\chi}_\Delta
,K]\chi_nh_\alpha\psi\right\|=O(\lambda n^{1/2})\|\chi_n \psi_{\alpha,\nu}\|.
\label{l10}
\end{equation}
Next, we estimate the first term on the r.h.s. of \fer{l8}:
\begin{eqnarray}
\|\overline{\chi}_\Delta\chi_nh_\alpha\psi\|&=&\|(L-e)^{-1}\overline{\chi}_\Delta (L-e)\chi_nh_\alpha\psi\|\nonumber\\
&\leq & C\|(L-e)\chi_nh_\alpha\psi\|\nonumber\\
&\leq& C\| n^{-1}\lambda [N,I]\chi_n'h_\alpha\psi\|+O(\lambda n^{-3/2})+C\|\chi_n(L-e)h_\alpha\psi\|\nonumber\\
&\leq& C\lambda n^{-1/2}\|\chi_n' \psi_{\alpha,\nu}\|+O(\lambda n^{-3/2}+\alpha n).
\label{sharpe}
\end{eqnarray}
Combining this with \fer{l10} and \fer{l8}, we arrive at the estimate
\begin{equation}
\left|\av{R}_{\chi_n \psi_{\alpha,\nu}}\right|\leq
C\lambda^2\|\chi_n' \psi_{\alpha,\nu}\|\,\|\chi_n \psi_{\alpha,\nu}\|+O(\lambda^2n^{-1}+\lambda\alpha n^{3/2}).
\label{l11}
\end{equation}
\indent
There is one more term in \fer{l7} we have to estimate:
$\av{\overline{\chi}_\Delta K
  \overline{\chi}_\Delta}_{\chi_n \psi_{\alpha,\nu}}$. Since
$P_\Omega^\perp(N+\lambda\tilde{I})P_\Omega^\perp\geq 0$ and since $P_\Omega\tilde{I}P_\Omega=0$, we have the bound
$
K\geq P_\Omega^\perp\lambda\tilde{I}P_\Omega +\mbox{adj.}\geq -C\lambda,
$
which implies
\begin{equation}
\av{\overline{\chi}_\Delta K
  \overline{\chi}_\Delta}_{\chi_n\psi_{\alpha,\nu}}\geq
  -C\lambda\|\overline{\chi}_\Delta\chi_n \psi_{\alpha,\nu}\|^2.
\label{l12}
\end{equation}
Using \fer{l12} and \fer{l11}, we obtain from \fer{l7}
\begin{eqnarray}
\av{K}_{\chi_n \psi_{\alpha,\nu}}&\geq&\theta\|\chi_n \psi_{\alpha,\nu}\|^2-(\theta+C\lambda)\|\overline{\chi}_\Delta\chi_n
\psi_{\alpha,\nu}\|^2 -C\theta\delta_{e,0}\|P_{\Omega_{\beta,0}}\chi_\Delta\chi_n
\psi_{\alpha,\nu}\|^2\nonumber\\
&&-C\lambda^2\|\chi_n'\psi_{\alpha,\nu}\|\,\|\chi_n\psi_{\alpha,\nu}\|-\lambda^{19/50}O(\alpha n+\lambda n^{-1/2})\nonumber \\
&&-\lambda O(\alpha n^{3/2}+\lambda n^{-1}).
\label{l13}
\end{eqnarray}
Next, we have for any $\eta,\epsilon>0$: 
\begin{eqnarray*}
\|\chi_n'\psi_{\alpha,\nu}\|\,\|\chi_n\psi_{\alpha,\nu}\|&\leq&\eta\|\chi_n\psi_{\alpha,\nu}\|^2+\eta^{-1}\|\chi_n'\psi_{\alpha,\nu}\|^2\\
&\leq&(\epsilon\eta^{-1}+\eta)\|\chi_n\psi_{\alpha,\nu}\|^2
+\eta^{-1}\epsilon^{-2}\|\overline{\chi}_n \psi_{\alpha,\nu}\|^2.
\end{eqnarray*}
In the second step, we used the standard fact that we can choose the partition of unity s.t. $\|\chi'_n\psi\|^2\leq\epsilon\|\chi_n\psi\|^2+\epsilon^{-2}\|\overline{\chi}_n\psi\|^2$, for any $\epsilon>0$. 
Combining this with \fer{l13}, we obtain from \fer{l6}:
\begin{eqnarray*}
\av{K}_{\psi_{\alpha,\nu}}&\geq&(\theta-C\lambda^2(\epsilon\eta^{-1}+\eta))\|\chi_n\psi_{\alpha,\nu}\|^2 +(n/2-C\lambda^2\eta^{-1}\epsilon^{-2})\|\overline{\chi}_n\psi_{\alpha,\nu}\|^2\\
&&-C\theta\delta_{e,0}\|P_{\Omega_{\beta,0}}\chi_\Delta\chi_n
\psi_{\alpha,\nu}\|^2 -(\theta+C\lambda)\|\overline{\chi}_\Delta\chi_n \psi_{\alpha,\nu}\|^2\\
&&-O(\lambda\alpha n^{3/2}+\lambda^{19/50}\alpha n+\lambda^{69/50}n^{-1/2}).
\end{eqnarray*}
Consider $\lambda$ small and fixed. Then if 
\begin{equation}
\frac{n}{2}-C\eta^{-1}\epsilon^{-2}\geq\theta,
\label{l13.1}
\end{equation}
we obtain
\begin{eqnarray}
\av{K}_{h_\alpha\psi_\nu}&\geq&
\theta\|h_\alpha\psi_\nu\|^2-C\theta\delta_{e,0}\|P_{\Omega_{\beta,0}}\chi_\Delta\chi_n
h_\alpha\psi_\nu\|^2\nonumber -O(\epsilon\eta^{-1}+\eta+\alpha n^{3/2}+n^{-1/2})\nonumber\\
&& -C\theta (n^{-1}+n^{-3}+\alpha^2n^2).
\label{l14}
\end{eqnarray}
On the last line, we used \fer{sharpe}. Let us choose the parameters as follows:
\begin{equation*}
\epsilon=\alpha^{1/10}, \ \eta=\alpha^{1/20}, \ n=\alpha^{-1/2}, 
\end{equation*}
then \fer{l13.1} is verified, and furthermore, \fer{l14} reduces to 
\begin{equation}
\av{K}_{\psi_{\alpha,\nu}}\geq\theta\|\psi_{\alpha,\nu}\|^2 -C\theta\delta_{e,0}\|P_{\Omega_{\beta,0}}\chi_\Delta\chi_n
\psi_{\alpha,\nu}\|^2-O(\alpha^{1/20}).
\label{l15}
\end{equation}
On the other hand, recalling \fer{l5}, we obtain by  choosing the parameters $\nu$ and $\alpha$ as $\nu=\alpha^3$:
\begin{equation}
\av{K}_{\psi_{\alpha,\nu}}\leq C\alpha^{1/2}.
\label{l16}
\end{equation}
Since $\|\psi_{\alpha,\nu}\|\rightarrow \|\psi\|=1$ as $\alpha,\nu\rightarrow
0$, and since
\begin{equation*}
-C\theta\delta_{e,0}\|P_{\Omega_{\beta,0}}\chi_\Delta\chi_n
\psi_{\alpha,\nu}\|^2\rightarrow-C\theta\delta_{e,0}\|P_{\Omega_{\beta,0}}P^\perp_{\Omega_{\beta,\lambda}}\psi\|^2
\end{equation*}
(recall that $\psi =P^\perp_{\Omega_{\beta,\lambda}}\psi$ if $e=0$), we obtain thus for small $\alpha$ from \fer{l15} and \fer{l16} the inequality
\begin{equation}
\frac{\theta}{2}\left(1-C\delta_{e,0}\|P_{\Omega_{\beta,0}}P^\perp_{\Omega_{\beta,\lambda}}\psi\|^2\right)\leq C\alpha^{1/2}.
\label{l17}
\end{equation}
For $e\neq 0$, this is a contradiction, and it shows that there can not  be
any eigenvalues of $L$ in the interval $\Delta$. Remark that there is no smallness
condition on the size of $\Delta$, except that it must not contain more than
one eigenvalue of $L_0$, so we can choose $\Delta=(e_-,e_+)$.\\
\indent
Let us look now at the case $e=0$. Again, we reach a contradiction from
\fer{l17}, provided
$
\|P_{\Omega_{\beta,0}}P^\perp_{\Omega_{\beta,\lambda}}\psi\|^2\lless 1.
$
 In this
case, we conclude that zero is a simple eigenvalue of $L$. Now the fact that $\|P_{\Omega_{\beta,0}}P^\perp_{\Omega_{\beta,\lambda}}\|=O(\beta|\lambda|)$ follows immediately from \fer{pertkms}, so
 taking $\beta|\lambda|$ small enough finishes the proof of
Theorem 2.2.\hfill $\blacksquare$
\\
\ \\

\centerline{\bf Acknowledgements}
The author thanks I.M. Sigal for his support and advice. Many thanks also go to J. Fr\"ohlich and R. Froese for stimulating discussions and to the referee for helpful comments. During the writing up of this work, the author has been supported by an NSERC PDF (Natural Sciences and Engineering Council of Canada Postdoctoral Fellowship), which is gratefully acknowledged.

\appendix

\section{Appendix}

\subsection{Selfadjointness of $L$ and some relative bounds}
We introduce  the positive operator
$
\Lambda={\rm d}\Gamma(|u|)
$
with domain $\dom(\Lambda)=\{\psi\in{\cal H}: \|\Lambda\psi\|<\infty\}$ and the number operator 
\begin{equation}
N={\rm d}\Gamma(1)
\label{nop}
\end{equation}
with natural domain $\dom(N)=\{\psi\in{\cal H}: \|N\psi\|<\infty\}$. \\

{\bf Proposition A.1\ (Relative Bounds).\ }{\it Set $L^2=L^2({\mathbb R}\times S^2)$, and let $0<\beta_0<\infty$ be a fixed number.
\begin{itemize}
\item[{\bf 1)}] If $f\in L^2$, then $||a(f)N^{-1/2}||\leq ||f||_{L^2}$.
 \item[{\bf 2)}] If $|u|^{-1/2}f\in L^2$, then
 $||a(f)\Lambda^{-1/2}||\leq ||\,|u|^{-1/2}f||_{L^2}$.
\item[{\bf 3)}] For $\psi\in\dom(N^{1/2})$ and $\psi\in\dom(\Lambda^{1/2})$ respectively, we have the following bounds, uniformly in $\beta\geq \beta_0$:
\begin{eqnarray*}
||I\psi||^2&\leq& C||G||\left( \|N^{1/2}\psi\|^2+\|\psi\|^2\right),\\
||I\psi||^2&\leq& C||G||\left( \|\Lambda^{1/2}\psi\|^2+\|\psi\|^2\right).
\end{eqnarray*}
Here, $C\leq C'(1+\beta_0^{-1})$, where $C'$ is independent of $\beta, \beta_0$.
\item[{\bf 4)}] For $\psi\in{\cal D}(N^{1/2})$, any $c>0$, and uniformly in $\beta\geq \beta_0$, we have
 \vspace*{-.1cm}
\begin{equation*}
\left|\scalprod{\psi}{\lambda I\psi}\right|\leq
c ||N^{1/2}\psi||^2+\frac{16\lambda^2}{c}||G||^2||\psi||^2 \int_{{\mathbb
R}^3}(1+\beta_0^{-1}\omega^{-1})|g|^2d^3k.
\end{equation*}
 \vspace*{-.75cm}
\item[{\bf 5)}] For $\psi\in{\cal D}(\Lambda^{1/2})$, any $c>0$, and
uniformly in $\beta\geq \beta_0$, we have
 \vspace*{-.1cm}
\begin{equation*}
\left|\scalprod{\psi}{\lambda I\psi}\right|\leq
c||\Lambda^{1/2}\psi||^2+\frac{32\lambda^2}{c}||G||^2||\psi||^2 \int_{{\mathbb
R}^3}(1+\beta_0^{-1}\omega^{-1})\frac{|g|^2}{\omega}d^3k.
\end{equation*}
\end{itemize}
}
{\it Remarks.\ } 1.\ The parameter $\beta_0$ gives the highest temperature, $T_0=1/\beta_0$, s.t. our estimates 3)-5) are valid uniformly in  $T\leq T_0$. $T_0$ can be fixed at any arbitrary large value. Since we are not interested in the large temperature limit $T\rightarrow\infty$, we set from now on for notational convenience $T_0=1$.\\
2.\ 
Notice that 4) and 5) tell us that $\forall c>0$ (with the $O$-notation introduced after Theorem 2.1),
\begin{equation*}
|\lambda I|\leq cN +O(\lambda^2/c),\ \ \ 
|\lambda I|\leq c\Lambda +O(\lambda^2/c),
\end{equation*}
where we understand these inequalities holding in a sense of quadratic forms
on $\dom(N^{1/2})$ and $\dom(\Lambda^{1/2})$ respectively.\\

{\it Proof of Proposition A.1.\ }
The proof is standard (see e.g. [BFS4], [JP1,2]), we only present the proof of 3), as an example of how to keep track of $\beta$.\\
\indent
From
$
\|I\psi\|^2\leq 4\|G\|^2\left( \|a^*(g_1)\psi\|^2+\|a^*(g_2)\psi\|^2+\|a(g_1)\psi\|^2+\|a(g_2)\psi\|^2\right),
$
and using the CCR $[a^*(f),a(g)]=\scalprod{f}{g}$, we get
$$
\|a^*(g_{1,2})\psi\|^2=\scalprod{\psi}{a(g_{1,2})a^*(g_{1,2})\psi}=\|a(g_{1,2})\psi\|^2+\|g_{1,2}\|^2_{L^2}\|\psi\|^2,
$$
so
$
\|I\psi\|^2\leq 8\|G\|^2\left( \|a(g_1)\psi\|^2+\|a(g_2)\|^2 +2\|g_1\|^2_{L^2}\|\psi\|^2\right),
$
where we used $\|g_1\|_{L^2}=\|g_2\|_{L^2}$, since $g_1(u,\alpha)=-g_2(-u,\alpha)$. Using 1) and 2) above, we get
\begin{eqnarray*}
\|I\psi\|^2&\leq& 16 \|G\|^2\|g_1\|^2_{L^2}\left( \|N^{1/2}\psi\|^2+\|\psi\|^2\right),\\
\|I\psi\|^2&\leq& 16 \|G\|^2\left\| |u|^{-1/2}g_1\right\|^2_{L^2}\left( \|\Lambda^{1/2}\psi\|^2+\|\psi\|^2\right).
\end{eqnarray*}
Next, we show that $\|g_1\|_{L^2}\leq C$ and $\|\,|u|^{-1/2}g_1\|_{L^2}\leq C$, uniformly in $\beta\geq \beta_0$. Indeed, notice that
$
\|g_1\|_{L^2}^2=\int_{{\mathbb R}^3}(1+2\mu)|g(\omega,\alpha)|^2d\omega dS(\alpha)=\|g_2\|_{L^2}^2,
$
where we represented $g$ in the integral in spherical coordinates. Since we
have $1+2\mu=1+2(e^{\beta\omega}-1)^{-1}\leq 1+2\beta^{-1}\omega^{-1}\leq
1+2\beta_0^{-1}\omega^{-1}$, uniformly in $\beta\geq \beta_0$, we get with \fer{IRUV} (for $p>0$) the following uniform bound in $\beta\geq \beta_0$:
\begin{equation}
\|g_{1,2}\|_{L^2}^2\leq 2\int_{{\mathbb R}^3}(1+\beta_0^{-1}\omega^{-1})|g(k)|^2d^3k =C<\infty.
\label{*}
\end{equation}
Similarly,
$
\|\,|u|^{-1/2}g_1\|_{L^2}^2\leq2\int_{{\mathbb R}^3}(1+\beta_0^{-1}\omega^{-1})\omega^{-1} |g(\omega,\alpha)|^2 d^3k=C<\infty,
$
uniformly in $\beta\geq \beta_0$. It is clear from the last two estimates that $C$ satisfies the bound indicated in the proposition. \hfill $\blacksquare$\\

These relative bounds and Nelson's commutator theorem yield  essential 
selfadjointness of the Liouvillian:
\\

{\bf Theorem A.2  (Selfadjointness of the Liouvillian).\ }{\it Since $H_p$ is bounded
below, there is a  $C>0$ s.t. $H_p> -C$. Suppose that
$[G,H_p](H_p+C)^{-1/2}$ is bounded in the sense
that the quadratic form
$\psi\mapsto 2i\IM \scalprod{G\psi}{H_p\psi}$, defined on $\dom(H_p)$, is
represented by an operator denoted $[G,H_p]_{{\rm o}}$, s.t. $[G,H_p]_{{\rm
    o}}(H_p+C)^{-1/2}$ is bounded. Then $\forall \lambda\in{\mathbb R}$, $L$ is essentially selfadjoint on
\begin{equation*}
\dom_0:=\dom(H_p)\otimes\dom(H_p)\otimes\dom(\Lambda)\subset{\cal
  H}_p\otimes{\cal H}_p\otimes{\cal
F}(L^2({\mathbb R}\times S^2)).
\end{equation*}}

{\it Proof.\ }
The proof uses Nelson's commutator theorem (see [RS],
Theorem X.37). Let 
$
{\cal N}=(H_p+C)\otimes{\bbbone}_p +\bbbone_p\otimes (H_p+C)+\Lambda +1,
$
then ${\cal N}$ is selfadjoint on $\dom_0$ and ${\cal N}\geq 1$. Also, $L$ is
defined and symmetric on $\dom_0$.\\
\indent
According to Nelson's commutator theorem, in order to prove Theorem 1.2, we have to show
that $\forall \psi\in\dom_0$ and some constant $d>0$,
\begin{eqnarray}
||L\psi||&\leq& d||{\cal N}\psi||,  \label{1}
\\
\left|\scalprod{L\psi}{{\cal N}\psi}-\scalprod{{\cal N}\psi}{L\psi}\right|&\leq& d||{\cal
N}^{1/2}\psi||^2.   \label{2}
\end{eqnarray}
Estimate \fer{1} easily follows from  $||L_p{\cal N}^{-1}||\leq 1$, $||L_f{\cal
N}^{-1}||\leq 1$ and $||I{\cal N}^{-1}||\leq||I(\Lambda+1)^{-1/2}||\,||(\Lambda+1)^{1/2}(\Lambda
+1)^{-1}||\leq d$ (by 3) of Proposition 6.1).
\\
\indent To show \fer{2}, notice that $L_0$ commutes with ${\cal N}$, so the l.h.s. of
\fer{2} reduces to
\begin{equation}
\left|\scalprod{I\psi}{{\cal N}\psi}-\scalprod{{\cal
N}\psi}{I\psi}\right| \leq \left|\scalprod{I\psi}{\Lambda\psi}-\scalprod{\Lambda\psi}{I\psi}\right| +K,\label{3}
\end{equation}
where 
\begin{eqnarray}
K&=&\left|\scalprod{I\psi}{((H_p+C)\otimes\bbbone+\bbbone\otimes(
    H_p+C))\psi}\right.\nonumber\\
& &\left. -\scalprod{((H_p+C)\otimes\bbbone+\bbbone\otimes( H_p+C))
\psi}{I\psi}\right|. \label{4}
\end{eqnarray}
Let us examine the first term on the r.h.s. of \fer{3}. It is easily shown that since $|u|g_{1,2}\in L^2({\mathbb R}\times S^2)$, then
$a^*(g_{1,2})\Lambda=\Lambda a^*(g_{1,2}) +a^*(|u|g_{1,2})$ on $\dom(\Lambda)$. This
shows that $a^{\#}(g_{1,2})$ leave $\dom(\Lambda)$ invariant and so we have $\forall \psi \in \dom_0$:
\begin{eqnarray*}
\lefteqn{ \left|\scalprod{I\psi}{\Lambda\psi}-\scalprod{\Lambda\psi}{I\psi}\right|}
\\
&=&\left|\scalprod{\psi}{(I\Lambda-\Lambda I)\psi}\right|\\
&=&\left|\scalprod{\psi}{\big( G_l\otimes ( a^*(|u|g_1)-a(|u|g_1)) -G_r\otimes
(a^*(|u|g_2)-a(|u|g_2))\big)\psi}\right|
\\
&\leq& c||\psi||\,||(\Lambda+1)^{1/2}\psi|| \leq c||{\cal N}^{1/2}\psi||^2,
\end{eqnarray*}
where we used Proposition 6.1 in the third step. \\
\indent
Now we look at $K$ given in \fer{4}. Using the specific form of $I$ (see
\fer{017}), we can write
$
K\leq |K_1|+|K_2|,
$
where
\begin{eqnarray*}
K_1&=&\scalprod{G_l\otimes(a(g_1)+a^*(g_1))\psi}{(H_p+C)\otimes\bbbone\psi}\\
&& -\scalprod{(H_p+C)\otimes\bbbone\psi}{G_l\otimes(a(g_1)+a^*(g_1))\psi},\\
K_2&=&\scalprod{G_r\otimes(a(g_2)+a^*(g_2))\psi}{\bbbone\otimes(H_p+C)\psi}\\
&& -\scalprod{\bbbone\otimes(H_p+C)\psi}{G_r\otimes(a(g_2)+a^*(g_2))\psi}.
\end{eqnarray*}
We examine $K_1$. Let
$\psi\in\dom_0$, then
$(H_p+C)^{1/2}\psi\in{\cal H}$, and so
\begin{eqnarray*}
K_1&=&2i\IM\scalprod{G_l\otimes(a(g_1)+a^*(g_1))}{(H_p+C)\otimes\bbbone\psi}\\
&=&2i\IM\scalprod{(a(g_1)+a^*(g_1))\psi}{[G,H_p]_o\psi},
\end{eqnarray*}
so we obtain
$
|K_1|\leq
 c\|(\Lambda+1)^{1/2}\psi\|\,\|(H_p+C)^{1/2}\otimes\bbbone\psi\| \leq c\|{\cal N}^{1/2}\psi\|^2.
$
The same estimate is obtained for $|K_2|$ in a similar way. This shows \fer{2} and completes the proof.\hfill $\blacksquare$

\subsection{Proof of Theorem 2.4}

For a fixed eigenvalue $e\neq 0$ of $L_0$, define the subsets of ${\mathbb N}$:
\begin{eqnarray*}
\nr^{(i)}&:=&\{j| E_i-E_j=e\},\\
\nl^{(j)}&:=&\{i| E_i-E_j=e\},\\
\nr &:=&\cup_{i}\ \nr^{(i)}=\{ j| E_i-E_j=e\mbox{\ for some $i$}\},\\
\nl &:=&\cup_{j}\ \nl^{(j)}=\{ i| E_i-E_j=e\mbox{\ for some $j$}\}.
\end{eqnarray*}
We also let $P_i$ denote the rank-one projector onto ${\mathbb C}\varphi_i$, where we recall that $\{\varphi_i\}$ is the orthonormal basis diagonalizing $H_p$. For any nonempty subset ${\cal N}\subset{\mathbb N}$, put 
\begin{equation*}
P_{{\cal N}}:=\sum_{j\in{\cal N}} P_j,\mbox{\ \ \ and\ \ \ }P_{\cal N}:=0 \mbox{\ if
  ${\cal N}$ is empty.}
\end{equation*}
Set $E_{mn}:=E_m-E_n$, and for $e\in\sigma(L_p)\backslash \{0\}$, $m\in{\cal N}_l$ and $n\in {\cal N}_r$, define:
\begin{eqnarray}
\delta_m&:=&\inf\sigma\Big(P_{\nr^{(m)}} GP_{{\cal N}_r^c} G
P_{\nr^{(m)}}\upharpoonright P_{\nr^{(m)}}   \Big)\geq 0,
\label{105}\\
\delta'_n&:=&\inf\sigma\Big(P_{\nl^{(n)}} GP_{{\cal N}_l^c}G P_{\nl^{(n)}}
\upharpoonright  P_{\nl^{(n)}} \Big)\geq 0.
\label{106}
\end{eqnarray}
Here, the superscript $\ ^{c}$ denotes the complement.
Notice that if $e=0$, then $\nr^c=\nl^c$ are empty, and $\delta_m,\delta_n'=0$. We define also $\delta_0:=\inf_{m\in\nl}\{\delta_m\}+\inf_{n\in\nr}\{\delta_n'\}$. From
$P(L_p=e)=\sum_{\{i,j:E_{ij}=e\}}P_i\otimes P_j$, we obtain together with the
definition of $\Gamma(e)$ given in \fer{Gamma}: 
\begin{equation}
\Gamma_p(e)=\sum_{m,n}\left(1-\delta_{E_{mn},e}\right)\sum_{\{i,j: E_{ij}=e\}}\sum_{\{k,l: E_{kl}=e\}}\int \delta(E_{mn}-e+u)P_{ij} \ m^* \ P_{mn}\  m\ P_{kl}.
\label{110}
\end{equation}
The idea here is to get a lower bound on the sum over $(m,n)\in{\mathbb
  N}\times{\mathbb N}$ by summing only over a convenient subset of ${\mathbb
  N}\times{\mathbb N}$ (notice that every term in the sum is positive). That
  subset is chosen such that the summands reduce to simpler expressions.\\
\indent
Using the definition of $m$ (see \fer{m}), we obtain
\begin{eqnarray*}
\lefteqn{
P_{ij}m^* P_{mn}m P_{kl}}\\
&=& P_{ij}\left(G_l\overline{g}_1 - G_r\overline{g}_2\right) P_{mn}\left( G_l g_1-G_r g_2\right) P_{kl}\\
&=& P_i GP_mGP_k\otimes P_n \delta_{jn}\delta_{nl} |g_1|^2 -P_i GP_m\otimes P_n {\cal C}G{\cal C}P_l \delta_{jn}\delta_{mk}\overline{g}_1 g_2\\
&& -P_m GP_k\otimes P_j {\cal C}G{\cal C}P_n \delta_{im}\delta_{nl} \overline{g}_2 g_1 +P_m\otimes P_j {\cal C}G{\cal C}P_n {\cal C}G{\cal C}P_l \delta_{im}\delta_{mk} |g_2|^2.
\end{eqnarray*}
Summing over $i,j$ and $k,l$ according to \fer{110} yields
\begin{eqnarray}
\lefteqn{\sum_{\{i,j: E_{ij}=e\}}\sum_{\{ k,l: E_{kl}=e\}} P_{ij} m^* P_{mn} m P_{kl}}\nonumber\\
&&=\left(\overline{g}_1 P_{{\cal N}_l^{(n)}} GP_m\otimes P_n -\overline{g}_2 P_m\otimes P_{{\cal N}_r^{(m)}} {\cal C}G{\cal C}P_n\right)\cdot\mbox{\ adjoint}.\nonumber
\end{eqnarray}
For $(m,n)\in{\cal N}_l\times{\cal N}_r^c$, we have $P_{{\cal N}_l^{(n)}}=0$
and $P_{{\cal N}_r^{(m)}}\neq 0$, and for $(m,n)\in {\cal N}_l^c\times{\cal
  N}_r$, we have $P_{{\cal N}_l^{(n)}}\neq 0$ and $P_{{\cal N}_r^{(m)}}=0$. As
explained above, we now get a lower bound on the sum \fer{110} by summing only
over the disjoint union
\begin{equation*}
(m,n)\in {\cal N}_l\times{\cal N}_r^c\ \dot{\cup}\ {\cal N}_l^c\times{\cal
  N}_r.
\end{equation*} 
An easy calculation shows that 
\begin{eqnarray*}
\Gamma_p(e)&\geq&\inf_{E_{ij}\neq 0}\left(\int_{S^2}dS \left|g_2(E_{ij},\alpha)\right|^2\right)\sum_{m\in{\cal N}_l}P_m\otimes{\cal C} P_{{\cal N}_r^{(m)}} G \ P_{{\cal N}_r^c} \ G P_{{\cal N}_r^{(m)}}{\cal C}\\
&&+\inf_{E_{ij}\neq 0}\left(\int_{S^2}dS \left|g_1(E_{ij},\alpha)\right|^2\right)\sum_{n\in{\cal N}_r} P_{{\cal N}_l^{(n)}} G \ P_{{\cal N}_l^c} \ G P_{{\cal N}_l^{(n)}}\otimes P_n.
\end{eqnarray*}
Next, we investigate the integrals. From \fer{i)}, we have 
\begin{equation*}
\int_{S^2}dS |g_{1,2}(E_{ij},\alpha)|^2 \geq |E_{ij}| \int_{S^2}dS |g(|E_{ij}|,\alpha)|^2,
\end{equation*}
uniformly in $\beta\geq 1$. With \fer{105}, \fer{106} and remarking that $\sigma({\cal C}T{\cal C})=\sigma(T)$ for any selfadjoint $T$, this yields
\begin{equation*}
\Gamma_p(e)\geq\inf_{E_{ij}\neq 0}\left( |E_{ij}|\int_{S^2}dS |g(E_{ij},\alpha)|^2\right)\left(\inf_{m\in{\cal N}_l}\{\delta_m\} +\inf_{n\in{\cal N}_r}\{\delta'_n\}\right) P(L_p=e),
\end{equation*}
since $\sum_{m\in{\cal N}_l} P_m\otimes P_{{\cal N}_r^{(m)}}=\sum_{n\in{\cal
    N}_r} P_{{\cal N}_l^{(n)}}\otimes P_n=P(L_p=e)$. This shows 1) of Theorem 4.4.\\
\indent
Now we look at the zero eigenvalue. A general normalized element of $\ran P(L_p=0)$ is of the form $\phi=\sum_ic_i\varphi_i\otimes\varphi_i$, with $\sum_i|c_i|^2=1$, so 
\begin{equation*}
\scalprod{\phi}{\Gamma(0)\phi}=\sum_{m,n}\left( 1-\delta_{E_{mn},0}\right)\sum_{i,j}\overline{c}_i c_j \int\delta(E_{mn}+u)\scalprod{\varphi_i\otimes\varphi_i}{m^* P_{mn} m\varphi_j\otimes\varphi_j}.
\end{equation*}
Using again the explicit form of $m$ given in \fer{m} and  $\scalprod{\varphi_m}{{\cal C}G{\cal C}\varphi_n}=\overline{\scalprod{\varphi_m}{G\varphi_n}}$, we obtain 
\begin{equation}
\scalprod{\phi}{\Gamma(0)\phi}=\sum_{m,n}\left(1-\delta_{E_{mn},0}\right)\int \delta(E_{mn}+u)\left|\scalprod{\varphi_n}{G\varphi_m}\right|^2 |c_ng_1-c_mg_2|^2.
\label{114}
\end{equation}
We split the domain of integration ${\mathbb R}\times S^2$ into ${\mathbb
  R}_+\times S^2\ \dot{\cup}\ {\mathbb R}_-\times S^2$ and using \fer{i)} and $g_2(u,\alpha)=-g_1(-u,\alpha)$, arrive at 
\begin{eqnarray*}
\lefteqn{\int \delta(E_{mn}+u) |c_ng_1-c_mg_2|^2}\\
&=&\int_{{\mathbb R}^3}\left\{\delta(E_{mn}+\omega)\left|\sqrt{1+\mu}c_n g-\sqrt{\mu}c_m g\right|^2\right. + \delta(E_{mn}-\omega)\left.\left|\sqrt{\mu}c_ng -\sqrt{1+\mu}c_mg\right|^2\right\}.
\end{eqnarray*}
This together with \fer{114} gives
\begin{eqnarray}
\scalprod{\phi}{\Gamma(0)\phi}
&=&2\sum_{\{m,n: E_{mn}<0\}}\left|\scalprod{\varphi_n}{G\varphi_m}\right|^2\frac{e^{\beta E_n}}{e^{-\beta E_{mn}}-1} \nonumber\\
&&\ \ \ \times \left| e^{-\beta E_m/2}c_n-e^{-\beta E_n/2}c_m\right|^2\int \delta(E_{mn}+\omega)|g|^2,
\label{115}
\end{eqnarray}
where we used $\delta(E_{mn}+\omega)\mu=\delta(E_{mn}+\omega) (e^{-\beta
  E_{mn}}-1)^{-1}$. Equation \fer{115} shows that if we choose
$c_n=Z_p^{-1/2}e^{-\beta E_n/2}$, then each term in the sum is zero. Recall
  now that the particle Gibbs state is given by \fer{particlegibbs}, 
so $\scalprod{\Omega_\beta^p}{\Gamma(0)\Omega_\beta^p}=0$. Since $\Gamma(0)\geq0$, this implies that $\Omega_\beta^p$ is a zero eigenvector of $\Gamma(0)$.\\
\indent
Finally we show that there is a gap in the spectrum of $\Gamma(0)$ at
zero. Indeed, from \fer{115}, we get by the definition of $g_0$ (see statement
of Theorem  4.4):
\begin{eqnarray*}
\scalprod{\phi}{\Gamma(0)\phi}&\geq& 2g_0\sum_{\{m,n: E_{mn}<0\}}|e^{-\beta E_m/2}c_n-e^{-\beta E_n/2}c_m|^2\\
&&=g_0\sum_{m,n}|e^{-\beta E_m/2}c_n-e^{-\beta E_n/2}c_m|^2\\
&&=g_0\sum_{m,n}\left( e^{-\beta E_m}|c_n|^2 +e^{-\beta E_n}|c_m|^2-e^{-\beta(E_m+E_n)/2}(\overline{c}_nc_m+c_n\overline{c}_m)\right)\\
&&=g_0\Big(Z_p(\beta)+Z_p(\beta)-2\Big|\sum_m e^{-\beta E_m/2}c_m\Big|^2\Big)\\
&&=2g_0Z_p(\beta)\Big( 1-\Big|\scalprod{\Omega_\beta^p}{\phi}\Big|^2\Big),
\end{eqnarray*}
where we used $\sum_n|c_n|^2=1$. Therefore, we obtain on $\ran
P_{\Omega_\beta^p}^\perp$:\  $\Gamma(0)\geq 2g_0Z_p(\beta)$. This proves that if $g_0>0$, then we have a gap at zero and zero is a simple eigenvalue.\hfill $\blacksquare$\\

\subsection{Proof of Proposition 4.7}

We denote the spectrum of $L_p$ by $\sigma(L_p)=\{e_j\}$, where we include
multiplicities, i.e. for degenerate eigenvalues, we have $e_j=e_k$ for different $j\neq k$. Let $P_j$ denote the rank one
projector onto $\Span\{\varphi_i\}$, where $\varphi_j\in{\cal H}_p\otimes{\cal H}_p$ is the unique eigenvector
corresponding to $e_j$. Let $e$ be a fixed eigenvalue of $L_p$. Setting $m_j=P_jm$, we have
\begin{eqnarray}
\lefteqn{ \scalprod{\psi}{Q_1\int m^* \left((L_p-e+u)^2+\epsilon^2\right)^{-1} m\
Q_1\psi}}\nonumber \\
&=&\sum_{e_j\in\sigma(L_p)}\scalprod{\psi}{Q_1\int m^*_j m_j
((e_j-e+u)^2+\epsilon^2)^{-1} Q_1\psi}.
\label{16}
\end{eqnarray}
First, we estimate the term in the sum coming from $\{j:e_j=e\}$:
\begin{equation}
\sum_{\{e_j=e\}}\scalprod{\psi}{Q_1\int m^*_j m_j(u^2+\epsilon^2)^{-1}
Q_1\psi}\leq\sum_{\{e_j=e\}}\int u^{-2}||m_j Q_1\psi||^2.
\label{17}
\end{equation}
Now
\begin{equation*}
\sum_{\{e_j=e\}} ||m_j Q_1\psi||^2= ||P(L_p=e)(G_l g_1-G_rg_2)Q_1\psi||^2\leq 2||G||^2(|g_1|^2+|g_2|^2)||\psi||^2,
\end{equation*}
so 
$
\fer{17}\leq 2||G||^2||\psi||^2\left(||g_1/u||^2_{L^2}+||g_2/u||^2_{L^2}\right)= 4||G||^2
||g_1/u||^2_{L^2} ||\psi||^2. 
$
From our assumptions on $g$ (see \fer{IRUV}) and \fer{i)}, it is clear that $\|g_1/u\|_{L^2}=C<\infty$, uniformly in $\beta\geq 1$, and we conclude that 
\begin{equation}
\fer{17}\leq C \|\psi\|^2.
\label{18}
\end{equation}
Next, we estimate the sum of the terms in  \fer{16} with $e_j\neq e$ and write it as 
\begin{equation}
\sum_{e_j\neq e}\int_{\mathbb R} du((e_j-e+u)^2+\epsilon^2)^{-1} \tilde{m}_j(u,\psi),
\label{19}
\end{equation}
where we put $\tilde{m}_j(u,\psi)=\int_{S^2}dS ||m_j(u,\alpha) Q_1\psi||^2$. $\forall
\xi>0$, we have
\begin{eqnarray}
\lefteqn{
\sum_{e_j\neq e}\int_{\{|u-(e-e_j)|\geq\xi\}} du \ ((e_j-e+u)^2+\epsilon^2)^{-1}
\tilde{m}_j(u,\psi)}\nonumber\\
&\leq&\xi^{-2}\sum_{e_j\neq e}\int_{\mathbb R}du\ \tilde{m}_j(u,\psi)\nonumber\\
&\leq&\xi^{-2}\int||m(u,\alpha)Q_1\psi||^2 \leq 4\xi^{-2}||G||^2 ||g_1/u||^2_{L^2}||\psi||^2 \leq C\xi^{-2}\|\psi\|^2.
\label{20}
\end{eqnarray}
Next, with the changes of variables $y=u-(e-e_j)$, we arrive at
\begin{eqnarray}
\lefteqn{
\sum_{e_j\neq e}\int_{\{|u-(e-e_j)|\leq\xi\}}du\
((e_j-e+u)^2+\epsilon^2)^{-1}\tilde{m}_j(u,\psi)}\nonumber \\
&=&\left(\int_{-\xi}^\xi dy\ (y^2+\epsilon^2)^{-1}\right)\sum_{e_j\neq
e}\tilde{m}_j(e-e_j,\psi)\nonumber\\
& &+\int_{-\xi}^\xi dy\ (y^2+\epsilon^2)^{-1}\sum_{e_j\neq e}\big[
\tilde{m}_j(y+e-e_j,\psi)-\tilde{m}_j(e-e_j,\psi)\big].
\label{21}
\end{eqnarray}
The mean value theorem yields for the last sum:
\begin{equation}
y\ \partial_y|_{\tilde{y}\in(-\xi,\xi)}\sum_{e_j\neq e}\tilde{m}_j(y+e-e_j,\psi).
\label{22}
\end{equation}
 Now
\begin{eqnarray*}
\lefteqn{\partial_y\sum_{e_j\neq e}\tilde{m}_j(y+e-e_j,\psi)}\nonumber \\
&=& 2\sum_{e_j\neq e}\int_{S^2}dS\ \RE\scalprod{P_j(\partial_u m)(y+e-e_j,\alpha)
Q_1\psi}{P_jm(y+e-e_j,\alpha)Q_1\psi}.\nonumber
\end{eqnarray*}
Using the Schwarz inequality for sums, we bound the modulus of the r.h.s. from
above by
\begin{equation}
2\int_{S^2}dS\ \sqrt{\sum_{e_j\neq e}||P_j(\partial_u
m)(y+e-e_j,\alpha)Q_1\psi||^2}\sqrt{\sum_{e_j\neq e}||P_jm(y+e-e_j,\alpha)Q_1\psi||^2}.
\label{23}
\end{equation}
Now $m(y+e-e_j,\alpha)=G_lg_1(y+e-e_j,\alpha)-G_rg_2(y+e-e_j,\alpha)$, so
\begin{eqnarray}
\lefteqn{
||P_jm(y+e-e_j,\alpha)Q_1\psi||^2}\label{above}\\
&\leq&2 |g_1(y+e-e_j,\alpha)|^2||P_jG_lQ_1\psi||^2 +2
|g_2(y+e-e_j,\alpha)|^2||P_jG_rQ_1\psi||^2\nonumber.
\end{eqnarray}
We have to evaluate this at $y=\tilde{y}\in(-\xi,\xi)$. Clearly,
$|e-e_j+\tilde{y}|\geq |e-e_j|-|\tilde{y}|>d_0-\xi\geq d_0/2$, if we choose
$\xi\leq d_0/2$, where 
\begin{equation*}
d_0:=\inf_{e_i\neq e_j}|e_i-e_j|>0.
\end{equation*}
The r.h.s. of \fer{above}  can thus be estimated from above by 
\begin{equation*}
2\sup_{|u|>d_0/2}|g_1(u,\alpha)|^2||P_jG_lQ_1\psi||^2 + 2\sup_{|u|>d_0/2}|g_2(u,\alpha)|^2||P_jG_rQ_1\psi||^2,
\end{equation*}
hence we arrive at
\begin{equation}
|\fer{22}| \leq 32|y|\,||G||^2||\psi||^2\int_{S^2}dS\
\left(\sup_{|u|>d_0/2}|\partial_ug_1|+\sup_{|u|>d_0/2}|g_1|\right).
\label{24}
\end{equation}
Using the conditions \fer{IRUV} with $p>0$, one shows that the suprema are bounded, uniformly in $\beta\geq 1$, and so is $|g_1|$, thus \fer{24} gives
\begin{equation}
|\fer{22}|\leq C |y|\, ||\psi||^2.
\label{25}
\end{equation}
Remark that the constant here depends on $d_0$, $C\sim d_0^{p-1/2}$. This argument is valid for any $p$.
Going back to the second term on the r.h.s. of \fer{21}, we have shown:
\begin{eqnarray}
\lefteqn{\hspace*{-2cm}\left|\int_{-\xi}^\xi\frac{dy}{y^2+\epsilon^2}\sum_{e_j\neq
e}\big[\tilde{m}_j(y+e-e_j,\psi)-\tilde{m}_j(e-e_j,\psi)\big]\right|}\nonumber\\
&\leq&C ||\psi||^2\int_{-\xi}^\xi\frac{|y|}{y^2+\epsilon^2}dy \leq C \frac{|\xi|}{\epsilon}||\psi||^2.
\label{26}
\end{eqnarray}
Now we consider the first term on the r.h.s. of \fer{21}. We see that, as $\epsilon/\xi\rightarrow 0$,
\begin{equation*}
\int_{-\xi}^\xi\frac{dy}{y^2+\epsilon^2}=\frac{2}{\epsilon}\Arctan(\xi/\epsilon)
=\frac{2}{\epsilon}\left(\frac{\pi}{2}+o\left((\epsilon/\xi)^\eta\right)\right),
\end{equation*}
for any $0<\eta<1$. This simply follows from the fact that for any such
$\eta$, we have $\lim_{x\rightarrow \infty} x^\eta(\Arctan(x)-\pi/2)=0$. Also, 
\begin{equation*}
\sum_{e_j\neq e}\tilde{m}_j(e-e_j,\psi)=\int\scalprod{\psi}{Q_1 m^*\delta(u-e+L_p)P(L_p\neq e) mQ_1\psi}.
\end{equation*}
We conclude that \fer{19} is equal to
\begin{equation*}
\frac{\pi}{\epsilon}\Big\{(1-O(\epsilon/\xi))\int\scalprod{\psi}{Q_1
m^*\delta(u-e+L_p)P(L_p\neq e) m Q_1\psi}-O(\xi+\epsilon \xi^{-2})||\psi||^2\Big\}.
\end{equation*}
Choose e.g. $\xi=\epsilon^{1/4}$ and $\eta$ close to $1$, then the we arrive at
\begin{equation*}
\fer{19}=\frac{\pi}{\epsilon}\left\{\int\scalprod{\psi}{Q_1 m^*\delta(u-e+L_p) P(L_p\neq e) m
Q_1\psi} -O(\epsilon^{1/4})||\psi||^2\right\}.
\end{equation*}
This together with \fer{18} yields
\begin{eqnarray*}
\lefteqn{Q_1\int m^*((L_p-e+u)^2+\epsilon^2)^{-1}m Q_1}\\
&\geq& Q_1\frac{\pi}{\epsilon}\left\{\int m^*P(L_p\neq e)\delta(L_p-e+u) m
-O(\epsilon^{1/4})\right\} Q_1.\ \ \ \  \blacksquare
\end{eqnarray*}

\subsection{Operator calculus}
We outline an operator calculus for functions of selfadjoint operators, used extensively in this work. For a detailed exposition and more references, we refer to [HS3].\\
\indent
Let $f\in C_0^k({\mathbb R})$, $k\geq 2$, and define the compactly supported complex measure
$
d\tilde{f}(z)=-\frac{1}{2\pi}\left(\partial_x+i\partial_y\right)\tilde{f}(z) dxdy,
$
where $z=x+iy$ and $\tilde{f}$ is an almost analytic complex extension of $f$ in the sense that 
$
\left(\partial_x+i\partial_y\right)\tilde{f}(z)=0,\ \ \ z\in{\mathbb R}.
$
Then, for a selfadjoint operator $A$, one shows that
\begin{equation*}
f(A)=\int d\tilde{f}(z) (A-z)^{-1},
\end{equation*}
where the integral is absolutely convergent. Given $f$, one can construct explicitely an almost analytic extension $\tilde{f}$ supported in a complex neighbourhood of the support of $f$. One shows that for $p\leq k-2$, 
\begin{equation}
\int\left|d\tilde{f}(z)\right|\,|\IM z|^{-p-1}\leq C\sum_{j=0}^k \|f^{(j)}\|_{j-p-1},
\label{norms}
\end{equation}
where 
$
\|f\|_n=\int dx \langle x\rangle^n |f(x)|,
$
and $\langle x\rangle=(1+x^2)^{1/2}$. Furthermore, the derivatives of $f(A)$ are given by 
\begin{equation}
f^{(p)}(A)=p!\int d\tilde{f}(A) (A-z)^{-p-1}.
\label{deriv}
\end{equation}
We finish this outline by mentioning that these results extend by a limiting argument to functions $f$ that do not have compact support, as long as the norms in the r.h.s. of \fer{norms} are finite.

\end{document}